\documentclass[doublecol]{epl2} 
\usepackage{graphicx,amsmath,amssymb,color,ulem}
\usepackage{nicefrac}
\usepackage{multirow, makecell, ragged2e}
\usepackage[titletoc,title]{appendix}
\usepackage{bm}
\usepackage{float}
\usepackage[titletoc,title]{appendix}

\newcommand{\be}{\begin{equation}}
\newcommand{\ee}{\end{equation}}

\newcommand{\ba}{\begin{eqnarray}}
\newcommand{\ea}{\end{eqnarray}}

\renewcommand{\vec}[1]{\bm{#1}}
\def\beq{\begin{equation}}
\def\eeq{\end{equation}}

\title{Hall Effect for Dirac Electrons in Graphene Exposed to an Abrikosov 
Flux Lattice}

\author{Jonathan Schirmer\inst{1,*}, Ravi Kumar\inst{2,*}, Vivas Bagwe\inst{3},Pratap Raychaudhuri\inst{3},Takashi Taniguchi\inst{4},Kenji Watanabe\inst{4},C. -X. Liu\inst{1},Anindya Das\inst{2} \and J. K. Jain\inst{1}}
\shortauthor{}

\institute{                    
  \inst{1} Department of Physics, 104 Davey Lab, The Pennsylvania State University, University Park, Pennsylvania 16802\\
  \inst{2} Department of Physics, Indian Institute of Science, Bangalore 560012, India\\
  \inst{3} Tata Institute of Fundamental Research, Homi Bhabha Road, Colaba, Mumbai 400 005, India\\
  \inst{4} National Institute for Materials Science, Namiki 1-1, Ibaraki 305-0044, Japan
}
\pacs{nn.mm.xx}{First pacs description}

\abstract{
The proposals for realizing exotic particles through coupling of quantum Hall effect to superconductivity involve spatially non-uniform magnetic fields. As a step toward that goal, we study, both theoretically and experimentally, a system of Dirac electrons exposed to an Abrikosov flux lattice. We theoretically find that non-uniform magnetic field causes a carrier-density dependent reduction of the Hall conductivity. Our studies show that this reduction originates from a rather subtle effect: a levitation of the Berry curvature within Landau levels broadened by the non-uniform magnetic field. Experimentally, we measure the magneto-transport in a monolayer graphene-hexagonal boron nitride - niobium diselenide (NbSe$_2$) heterostructure, and find a density-dependent reduction of the Hall resistivity of graphene as the temperature is lowered from above the superconducting critical temperature of NbSe$_2$, when the magnetic field is uniform, to below, where the magnetic field bunches into an Abrikosov flux lattice.}

\begin{document}

\maketitle

\section{Introduction}
A promising strategy for creating particles obeying non-Abelian braiding statistics, which may be useful for quantum computation~\cite{Nayak08}, is to integrate superconductivity into the integer quantum Hall (QH) effect, which can produce topological superconductivity (TS) supporting Majorana particles \cite{Alicea12,Rickhaus12,Leijnse12,Stanescu13,Elliott15,Amet16,Shalom16,Alicea16,Lee17,Sahu18,Jeon19,Chaudhary20}.
One of the (many) practical challenges toward achieving this goal is that the strong magnetic field required for  QH effect usually kills superconductivity. Even if systems can be found wherein superconductivity survives to sufficiently high magnetic fields, a fundamental difficulty is that a gapped QH state does not couple to a superconductor in a meaningful manner. One can overcome this impediment by considering non-uniform magnetic fields, which produce ``dispersive" Landau levels and thus allow, when partially filled, proximity coupling to superconductivity \cite{Jeon19,Chaudhary20} 
This motivates, as a first step, the need to gain an understanding of QH effect in the presence of non-uniform magnetic fields.
In practice, such non-uniform fields are naturally produced by the nearby superconductor itself, in the form of an Abrikosov flux lattice.

We study, both theoretically and experimentally, a system of Dirac electrons exposed to an Abrikosov flux lattice. Theoretically, we find that non-uniformity of the magnetic field causes a carrier-density-dependent reduction of the Hall conductivity. We explain this reduction qualitatively and semi-quantitatively as a consequence of a rather subtle effect, namely levitation of the Berry curvature in broadened Landau levels. Experimentally, we explore magnetotransport in a monolayer graphene-hBN-NbSe$_2$ heterostructure, in which the electron and hole gases in graphene are exposed to an Abrikosov flux lattice threading through the NbSe$_2$ layer for temperatures below $T_c$. We contrast our results with an earlier work~\cite{Geim92} below.

\begin{figure}[t]
\includegraphics[width=.49\textwidth]{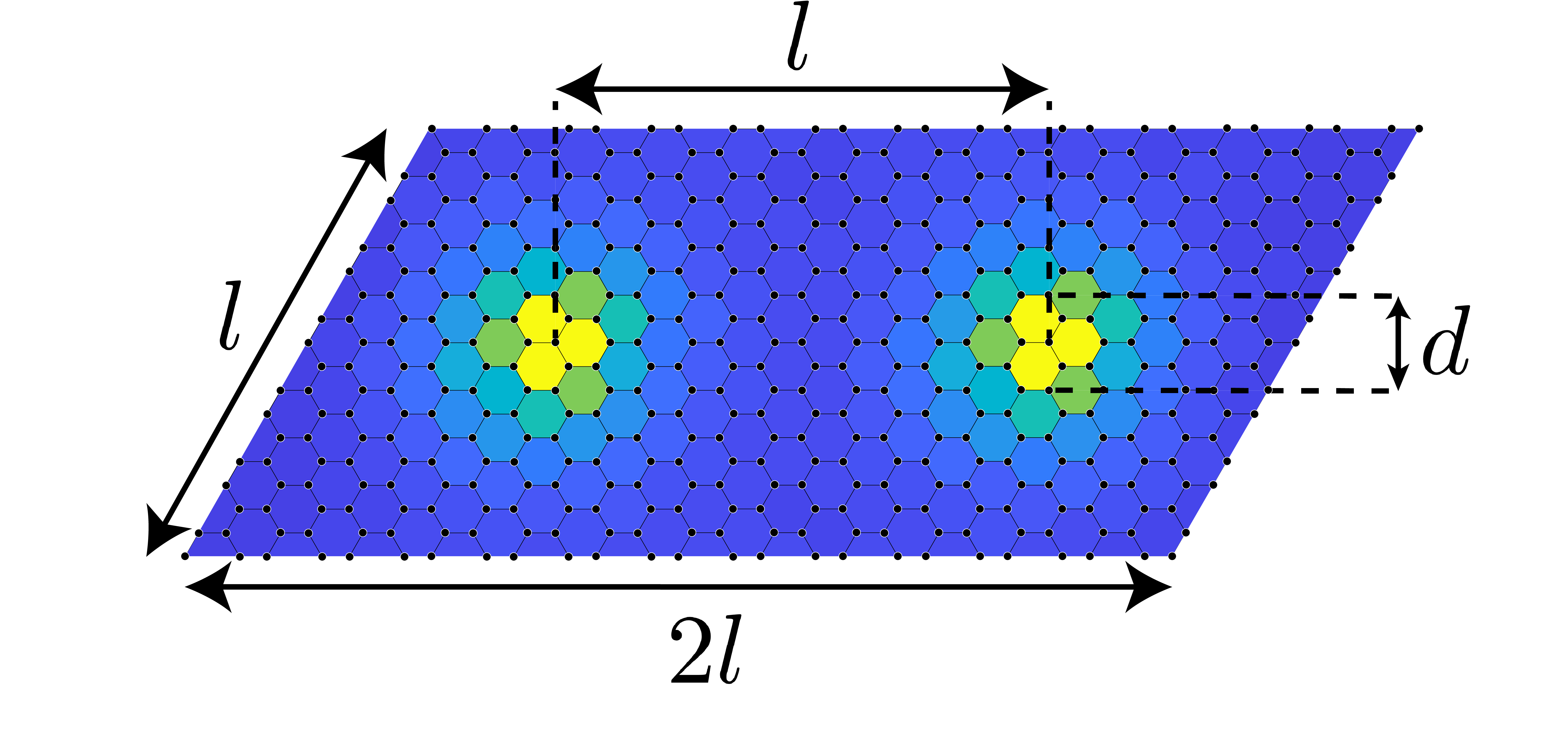}
\caption{The magnetic flux profile within in a magnetic unit cell.
The lattice constant of the triangular Abrikosov flux lattice, $l$, 
Each flux of size $d$ is centered on a site and is modeled with the Lorentzian profile of Eq.~\ref{magnetic_field}.}
\label{vortex}
\end{figure}

\begin{figure*}[t]
\includegraphics[width=\textwidth]{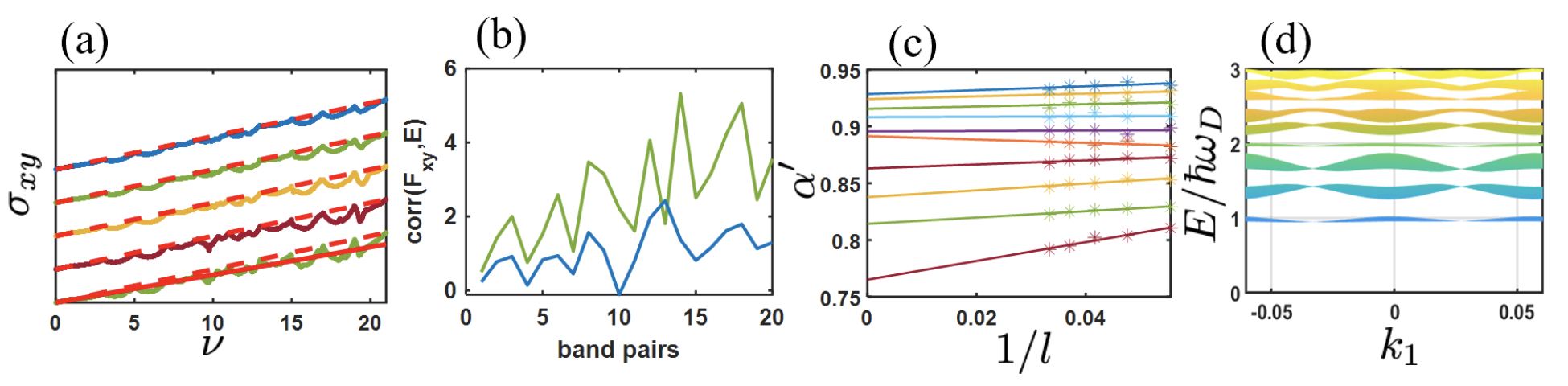}
\caption{(a) The colored curves show the Hall conductivity $\sigma_{\rm xy}$ (in units of $e^2/h$) plotted against filling factor $\nu$ for (from the bottom to top) $d/l=$ 0.04, 0.08, 0.12, 0.16, 0.2. All curves begin at $\sigma_{\rm xy}=0$ but are vertically offset for ease of comparison. The dashed red lines above each colored curve show the Hall conductivity in uniform field. The solid red line is the best linear fit (shown only for $d/l=0.04$ to avoid clutter), the slope of which gives $\alpha'$. (b) The correlation between the Berry curvature and energy (Eq. \ref{corr}) for pairs of degenerate bands up to pair 20. The green curve is for $d/l=$ 0.04 and the blue curve is form $d/l=$ 0.2. The correlation is generally positive for each pair and increases as $d/l$ is made smaller. (c) Continuum limit extrapolations of $\alpha'$  for (from bottom to top)  $d/l=0.02, 0.04, 0.06\ldots, 0.2$. The lines are linear fits. (d) The LL bands for $d/l=0.04$ as a function of $k_1$ as viewed along the $k_2$ direction (where $\hat{\vec{k}}_1=\frac{\sqrt{3}}{2}\hat{\vec{x}}-\frac{1}{2}\hat{\vec{y}}$, $\hat{\vec{k}}_2=\hat{\vec{y}}$ are magnetic reciprocal lattice unit vectors.) The bands at $E=0$ are flat and not visible in this figure.
Whenever the chemical potential lies in a gap between two successive bands, $\sigma_{\rm xy}$ touches its classical values, as seen in panel (a).
\label{wide_col}}
\end{figure*}

\section{Theory}
We consider the model of non-interacting electrons hopping on a lattice subject to an applied magnetic field. Because the Abrikosov vortices themselves form a triangular lattice, it is most natural to consider electrons hopping in a hexagonal lattice, for which we have chosen the honeycomb lattice, which produces Dirac electrons.
The Hamiltonian for this model is
\begin{equation} \label{eq5}
\hat{H}=-\sum_{\langle m,n \rangle}{te^{i\theta_{m,n}}c^\dagger_{m}c_{n}}
\end{equation}
where $c^\dagger_{m}$ and $c_m$ are creation and annihilation operators for electrons at site $m$, $te^{i\theta_{m,n}}$ is the hopping matrix element from site $n$ to $m$, the phases $\theta_{m,n}$ are determined by the local magnetic field up to a gauge transformation, and the sum $\langle m,n\rangle$ is over nearest neighbor sites.
 We denote the distance between neighboring sites by $a$ and quote all lengths in units of $a$. We choose the magnetic unit cell (MUC) so that the net flux through it is $\Phi_0 = h/e$. Our choice of periodic gauge is discussed in the SM \cite{SM-Schirmer20}. The energy spectrum of the honeycomb lattice near $E=0$ in uniform magnetic field consists of flat Landau bands which come in degenerate pairs. This degeneracy is due to the presence of two Dirac cones in the $B=0$ spectrum which occur at inequivalent momenta.  The spectrum is discussed in more detail in the SM \cite{SM-Schirmer20}.

Our neglect of interaction between electrons is justified for the following reasons. To begin with, we are concerned with high filling factors where correlation effects are not dominant. Further, the main effect of interaction is to renormalize the Fermi velocity of Dirac electrons, to which our results below are not sensitive. Finally, we focus here on the {\it changes} that arise when the temperature is reduced from above $T_c$ to below, and do not expect the effect of interaction to change substantially during this process.

We model the magnetic field of an Abrikosov vortex as
\begin{equation}
B(r)=\frac{C}{r^2+(d/2)^2}.
\label{magnetic_field}
\end{equation}
Here $d$ characterizes the size of the vortex, $r$ is the distance from the center, and the constant $C$ is chosen so that the total flux through each vortex (integrated over the MUC) is the superconducting flux quantum $\Phi_0/2$. The separation between two neighboring vortices will be denoted by $l$, which is also proportional to the lattice constants of the MUC, as shown in Fig.~\ref{vortex}.
We shall present our results in terms of the dimensionless parameter $d/l$, and will be interested in the continuum limit, where both $d$ and $l$ tend to infinity, with the ratio $d/l$ held constant.

To calculate the Hall conductivity using the method first described by Thouless {\it et al.} \cite{Thouless82}, one defines the non-Abelian Berry connection, determined by the eigenstates $|\Psi^n(\vec{k})\rangle = e^{i\vec{k}\cdot\vec{r}} |u^n(\vec{k})\rangle$, as
\begin{equation}
A^{mn}_{\mu}(\vec{k})=i\langle u^m(\vec{k})| \partial_{\mu} |u^n(\vec{k})\rangle
\end{equation}
where $\partial_{\mu}$ is the shorthand for $\partial/\partial{k_{\mu}}$. The Berry curvature $F$ is then defined as
\begin{equation}
 F_{\mu\nu}(\vec{k})=\partial_{\mu} A_{\nu} - \partial_{\nu} A_{\mu} + i \left[{A}_{\mu}, {A}_{\nu} \right]
\end{equation}
The only non-zero compnents of $F$ are $F_{\rm xy}=-F_{\rm yx}$. The Hall conductivity is then given, in units of $e^2/h$, by the Berry curvature integrated over the Brillouin zone:
\begin{equation} \label{hall_formula}
\sigma_{\rm xy} = \frac{1}{2\pi} \int_{\rm BZ} d^2k  {\rm Tr} \left[ { F}_{\rm xy}(\vec{k}) \right]_{E(\vec{k})<E_{\rm F}}
\end{equation}
where the trace is over occupied states. If the Fermi level $E_{\rm F}$ lies in a gap, $\sigma_{\rm xy}$ is given by a topological index called the first Chern number $C$, which takes integer values. If the Fermi level lies in a band, the Hall conductivity is no longer an integer. We compute Eq. \ref{hall_formula} numerically~\cite{Fukui05}; see SM for details \cite{SM-Schirmer20}.

 For values of $d/l$ corresponding to the experiment, we find that the computed values of $\sigma_{\rm xy}$ for non-uniform magnetic fields are less than those for uniform magnetic fields, as shown in Fig. \ref{wide_col}a. To quantify the reduction of Hall {\it conductivity}, we introduce the quantity
\begin{equation}
\alpha'=\frac{\sigma_{xy}(T<T_c)}{\sigma_{xy}(T>T_c)}.
\end{equation}
The ratio $\alpha'$ is more natural than $\alpha$ from the theoretical perspective.
As indicated in Fig.~\ref{wide_col}a, $\sigma_{\rm xy}$ varies linearly with $\nu$ for uniform $B$, as expected, and approximately linearly with $\nu$ also for non-uniform $B$ over the $\nu$ range considered in our calculation ($\nu\leq21$).  The reduction increases with decreasing $d/l$.
An earlier calculation \cite{Nielsen95} of $\alpha'$ did not evaluate the continuum limit, which, as we see below, significantly alters the results. We also use a more realistic profile for the flux, placing the center of the vortex at a lattice site, as opposed to at the center of a plaquette, which gives a much better convergence to the continuum limit for small $d/l$. Additionally, we work with Dirac electrons as appropriate for graphene. \par

An important question is: What is the physical mechanism for the reduction of Hall conductivity in the presence of non-uniform magnetic fields? An obvious possibility is that some of the bands (see Fig.~\ref{wide_col}d) become trivial, i.e. have $C=0$, and thus do not contribute to the Hall conductivity. However, we find in our calculations that all physically relevant degenerate pairs of Landau bands carry shared Chern number $C=2$ even for the most nonuniform magnetic field that we have considered ($d/l=0.02$).
Certain bands have large and negative $C$ (to ensure that the total Chern number is zero), but these bands occur at relatively high fillings (see SM for further details \cite{SM-Schirmer20}); we include only bands that are sufficiently far from these bands.

\begin{figure}[t]
	\includegraphics[width=0.45\textwidth]{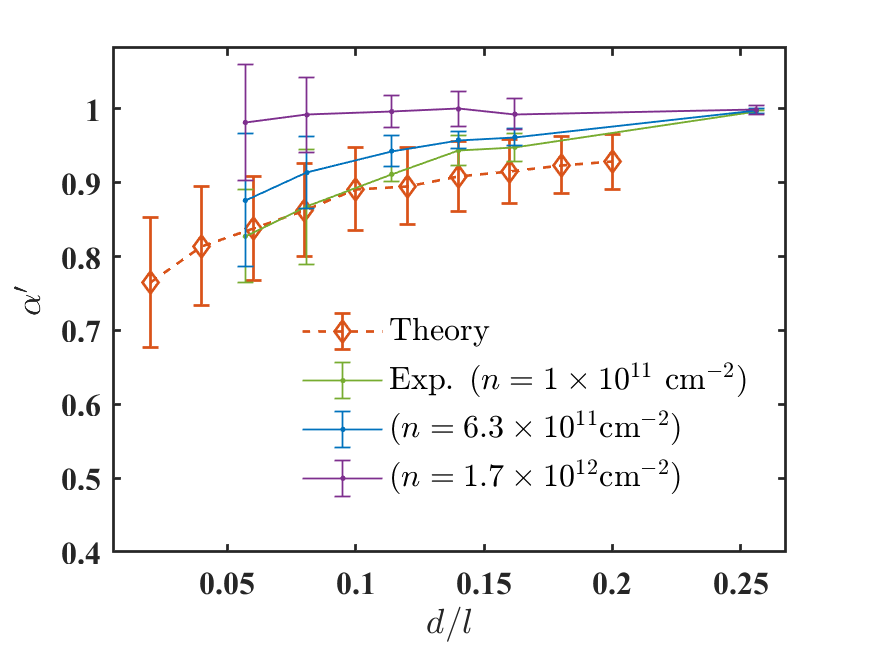}
	\caption{
  Comparison between experiment and theory for the $d/l$ dependence of $\alpha'$. The theoretical curve is obtained by averaging up to $\nu_{\rm max}=21$, which corresponds to a density of 6$\times$10$^{10}$ cm$^{-2}$ (6$\times$10$^{8}$ cm$^{-2}$) for $d/l=0.2$ ($d/l=0.02$).}
	\label{theory-exp}
\end{figure}

\begin{figure}
	\includegraphics[width=0.45\textwidth]{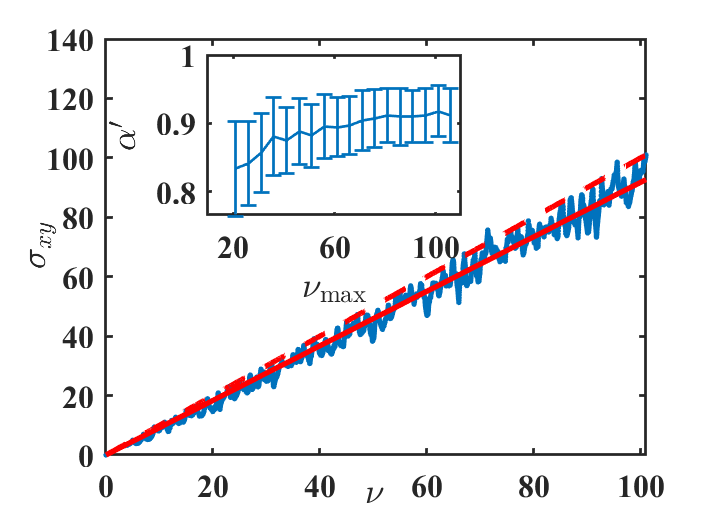}
	\caption{The Hall conductivity, along with the best linear fit up to $\nu=\nu_{\rm max}=106$, for a calculation with $l=30$ and $d/l=0.04$. The inset shows the dependence of $\alpha'$ on $\nu_{\rm max}$. The shown range of $\nu_{\rm max}$ corresponds to densities $2.4\times 10^{9}$ cm$^{-2}$ to  $1.2\times 10^{10}$ cm$^{-2}$. }
	\label{high_nu}
\end{figure}

The bands are broadened due to nonuniform $B$, but that also would not cause a correction to $\sigma_{\rm xy}$ so long as the Berry curvature is uniformly distributed in energy. This implies a subtle origin for the reduction of $\sigma_{\rm xy}$, namely a levitation of the Berry curvature within each broadened Landau band.
This is corroborated by computation (for degenerate pairs of bands) of the correlation between the Berry curvature $F_{\rm xy}(\vec{k})$ and energy $E(\vec{k})$, given by
\begin{equation}
{\rm corr}(F_{\rm xy},E)=\frac{1}{\Gamma} \int_{BZ} d^2k \big(E(\vec{k})-\bar{E} \big) F_{\rm xy}(\vec{k})
\label{corr}
\end{equation}
where $\Gamma$ is the bandwidth and $\bar{E}$ is the average energy of the band. We find that this quantity is generally positive (Fig. \ref{wide_col}b), 
implying that the Berry curvature tends to move away from the low energy states and concentrate near the high energy states.
When the Fermi energy enter into a band, the low energy states are occupied first and thus contribute less Berry curvature for non-uniform magnetic $B$.

Because both the magnetic length and size of the Abrikosov fluxes are large compared to the lattice constant for the experimental parameters, we focus on the continuum limit where $l\rightarrow \infty$ with $d/l$ constant. The extrapolation of $\alpha'$ to the continuum limit is shown in Fig. \ref{wide_col}c, where we have chosen to fit within the filling factor region $0\leq \nu\leq 21$.
The resulting continuum values of $\alpha'$ as a function of $d/l$ are shown by diamonds in Fig. \ref{theory-exp}. 
We have used the value of $d = 30$ nm~\cite{Galvis18}. The filling factor range $0\leq \nu\leq 21$ accessible to our theory corresponds to carrier concentration up to 6$\times$10$^{10}$ cm$^{-2}$. 

Fig.~\ref{high_nu} shows the reduction factor $\alpha'$ for $d/l=0.04$ when fitted up to $\nu_{\rm max}=106$; the value $\alpha'=0.92\pm0.04$ is larger than $\alpha'=0.83 \pm 0.07$ obtained by fitting up to $\nu_{\rm max}=21$ (Fig.~\ref{wide_col}a). The quantity $\alpha'$ thus
depends on the filling factor range available, and, in fact, increases as the filling factor increases, as shown in the inset of Fig.~\ref{high_nu}. We stress that the results shown in the inset are for $l=30$, and do not represent the continuum limit. The increase in $\alpha'$ can be understood from the observation that the reduction is roughly proportional to the fraction of Landau bands that cross the Fermi energy; this fraction vanishes in the limit of very high fillings. This feature of theory is also consistent with the experimental density dependence of $\alpha'$. We are not able to obtain continuum limit values of $\alpha'$ for large values of $\nu_{\rm max}$, but cannot rule out the possibility that $\alpha'\rightarrow 1$ in the limit of very large $\nu_{\rm max}$ or large densities (see SM for further details \cite{SM-Schirmer20}).
\section{Experiment}

\begin{figure}
\includegraphics[width=.485\textwidth]{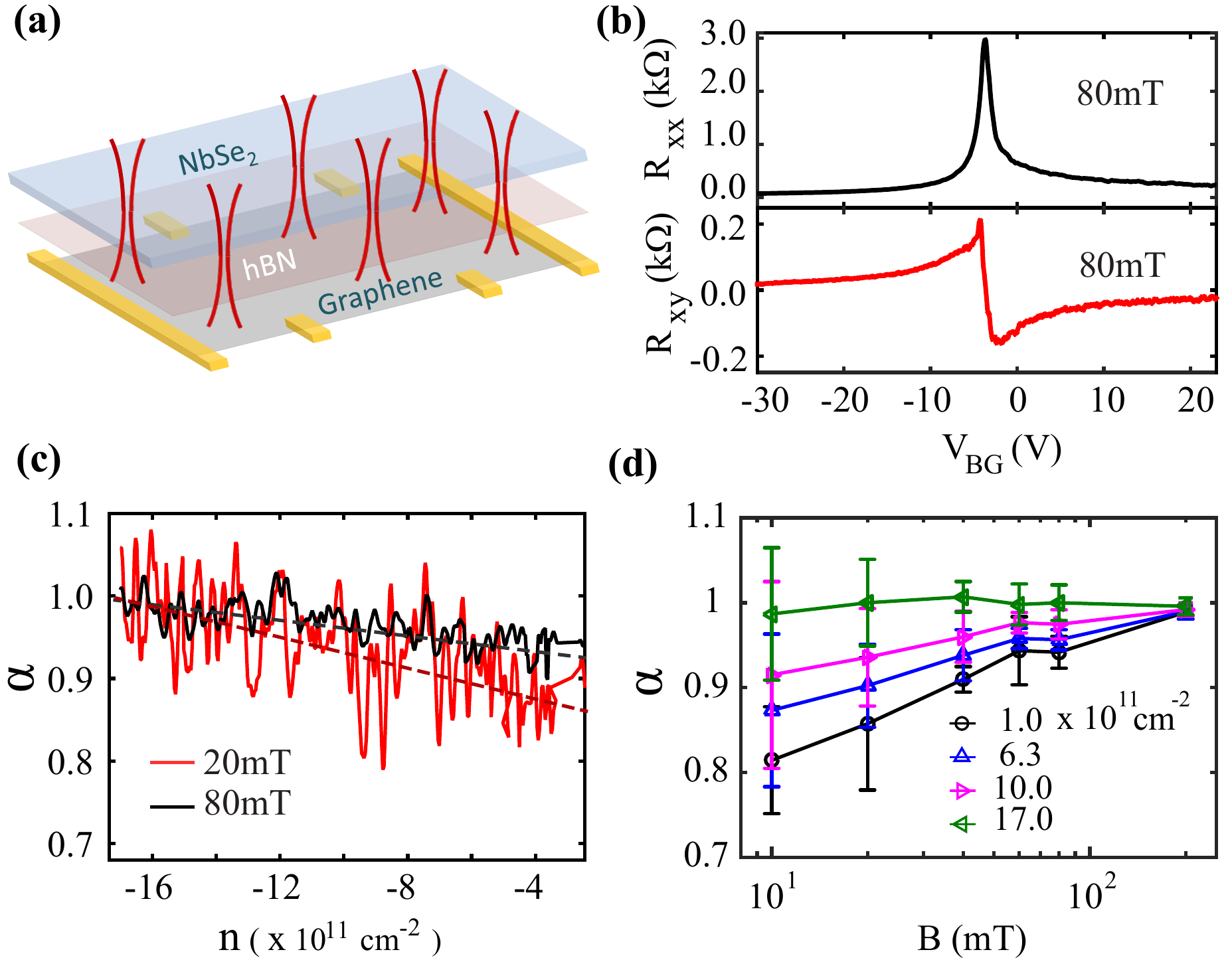}
\caption{(a) Device Schematic: Abrikosov vortices (red color) are formed in the $\sim$ 40 nm thick NbSe$_2$ flake and threaded into graphene layer residing at $\sim$ 10 nm (hBN thickness) below the NbSe$_2$. A Constant current is injected in the graphene channel using the left-most and right-most contacts. The longitudinal and transverse Hall voltages are measured using the inner four contacts. (b) Measured longitudinal and Hall resistance as a function of the back gate voltage at $B$ $=80mT$ and temperature $=250mK$.(c) $\alpha$ $\equiv R_{\rm xy}(T<T_{\rm c}) / R_{\rm xy}(T>T_{\rm c})$ for $20mT$ (solid red line) and $80mT$ (solid black line) as a function of hole carrier concentration. The dashed lines are the linear fittings to obtain the mean value. (d) Mean value of the $\alpha$ as a function of magnetic field for several values of $n$. The error bar indicates the standard deviation of the fluctuations of $\alpha$ in figure 1c.}
\label{schematic}
\end{figure}

In this section, we will present our experimental results. Device Schematic and measurement setup for the experiment are shown in Fig.\ref{schematic}a.  The device consists of graphene/hBN heterostructure on a SiO$_2$/Si substrate. The heterostructure was achieved by the standard hot pick-up technique \cite{Wang13}. Contacts were made on graphene in Hall bar geometry using E-beam lithography, and followed by thermal deposition of Cr(5 nm)/Pd (13 nm)/Au (70 nm). A top hBN then covered the device and followed by transfer of a thin NbSe$_2$ flake ($\sim$ 40nm) as highlighted in our previous work \cite{Sahu16}. The top hBN has a thickness of $\sim$ 10 nm, which acts as an insulating layer between the graphene channel and the NbSe$_2$ superconductor. As discussed in the Supplementary Material (SM)~\cite{SM-Schirmer20}, this thickness prevents any proximity induced superconductivity in graphene layer; we have also ruled out shorting between graphene and NbSe$_2$ from a lack of current between them at room temperature.
Standard four-probe measurements are performed using Lock-In amplifier by current biasing the graphene channel and measuring $V_{\rm xx}$ (longitudinal voltage) and $V_{\rm xy}$ (transverse voltage) simultaneously. The mobility of the graphene was $\sim$ 68,000 cm$^2$/V/s. The measurements were done in a He3 cryostat in the temperature (T) range of 250 mK to 15 K.
The measured longitudinal ($R_{\rm xx}$) and Hall resistance ($R_{\rm xy}$) as a function of the back gate voltage ($V_{\rm BG}$) at magnetic field $80~{\rm mT}$ and temperature $250~{\rm mK}$ are shown in Fig. \ref{schematic}b. The $R_{\rm xx}$ and $R_{\rm xy}$ were measured at several applied magnetic fields in the range from 10 mT to 200 mT. We have measured $R_{\rm xy}$ for +B and -B, then taken averaged $R_{\rm xy}$ in order to eliminate the finite contribution coming from the $R_{\rm xx}$ component (see SM\cite{SM-Schirmer20} for details). Similarly, we have measured the $R_{\rm xx}$ and $R_{\rm xy}$ at several magnetic fields above $T_c$ (see SM). Fig. \ref{schematic}c shows the ratio $\alpha \equiv R_{\rm xy}(T<T_{\rm c}) / R_{\rm xy}(T>T_{\rm c})$ as a function of the carrier concentration ($n$) for 20 mT and 80 mT. In Fig. \ref{schematic}c, the solid lines are the raw data and the dashed linear lines are the linear fittings to obtain the mean value. It can be seen from Fig. \ref{schematic}c that $\alpha$ fluctuates around unity at higher $n$ and gradually decreases with lowering $n$. The fluctuations in $\alpha$, which decrease with increasing magnetic field, are a mesoscopic phenomenon. In Fig. \ref{schematic}c we show the data for the hole side as the fluctuations for hole carriers are of smaller magnitude,
as can be seen in the $R_{\rm xy}$ plot in Fig. \ref{schematic}b. The carrier concentration is
determined by $ne = C_{\rm g}V_{\rm BG}$, where $C_{\rm g}$ is the geometrical capacitance of SiO$_{2}$ dielectric. In Fig. \ref{schematic}d, we summarize the mean value of $\alpha$ as a function of magnetic fields for several values of $n$.  Its value
decreases monotonically with $n$, and increases with magnetic field, reaching unity around B$\sim$200 mT. In order to compare with the theoretical results, one needs to convert the Hall resistivity to Hall conductivity. As can be seen from the SM that for our experiment, $R_{xx} > R_{xy}$, 
implying $R_{xx}^2 \gg R_{xy}^2$. As a result, 
$\sigma_{xy} \propto R_{xy}$. Thus, $\alpha' = \sigma_{ xy}(T<T_c) / \sigma_{ xy}(T>T_c)$ will be very similar to $\alpha \equiv R_{\rm xy}(T<T_{\rm c}) / R_{\rm xy}(T>T_{\rm c})$, as shown by the green line in Fig. \ref{theory-exp}. One can find a qualitative agreement between theoretical and esperimental results. 
\section{Discussions}
One may wonder if other physics might contribute to the reduction of the Hall conductivity, such as hydrodynamic \cite{Berdyugin19} or ballistic effects \cite{Beenakker89}. As discussed in detail in the SM, these effects are not relevant under the present experimental conditions: the hydrodynamic effects are seen only in extremely clean samples in the intermediate temperature regime (50-150K) and ballistic effects are seen only in narrow channels\cite{Beenakker89}. How these effects are modified for non-uniform magnetic fields is an interesting question for future studies. We note that a previous work~\cite{Geim92} investigated similar physics but in a different regime. They studied electrons with parabolic dispersion (GaAs based two-dimensional electron gas), and the locations of the fluxoids in that work were determined mainly by the presence of random disorder in the superconductor. While they also observed a reduction of the Hall resistance, they found that $\alpha$ saturates for very small $d/l$. The origin of this difference is not clear at the moment.

\section{Conclusions}
In conclusion, our principal finding is that a non-uniformity in magnetic field causes a levitation of the Berry curvature in broadened Landau levels, which, in turn, results in a reduction of the Hall conductivity, as seen qualitatively by our experiment. Another aspect of our calculations is worth stressing, that the low lying Landau levels remain surprisingly well defined even for highly non-uniform magnetic fields (see Fig.~\ref{wide_col}d), implying that QH effect ought to be observable even when the size of the Abrikosov fluxes is much smaller than their separation. This provides an optimistic outlook for the realization of topological superconductivity by coupling QH effect and superconductivity.


\acknowledgments
The work at Penn State was supported in part by the U. S. Department of Energy,
Office of Basic Energy Sciences, under Grant No. DE-SC0005042 (J.K.J.); 
by Kaufman New Initiative research grant KA2018-98553 of the Pittsburgh Foundation (J. S. and C.X.L.); and by the Office of Naval Research (Grant No. N00014-18-1-2793) and the U.S. Department of Energy (Grant No.~DESC0019064) (C.X.L.).
The numerical calculations were performed using Advanced CyberInfrastructure computational resources provided by The Institute for CyberScience at The Pennsylvania State University. AD thanks Ministry of human resource development (MHRD), India for the financial support under Scheme for Promotion of Academic and Research Collaboration (SPARC/2018-2019/P1178/SL).


\begin{thebibliography}{0}

\bibitem{Nayak08}
  \Name{Chetan Nayak, Steven H. Simon, Ady Stern, Michael Freedman, and Sankar Das Sarma}
  \REVIEW{Rev. Mod. Phys.} {80}{2008}{1083}
  
  \bibitem{Alicea12}
  \Name{J. Alicea}
  \REVIEW{Reports on Progress in Physics} {75}{2012}{076501}

  \bibitem{Rickhaus12}
  \Name{P. Rickhaus, M. Weiss, L. Marot, and C. Schonenberger}
  \REVIEW{Nano Letters} {12}{2012}{1942}
  
  \bibitem{Leijnse12}
  \Name{M. Leijnse and K. Flensberg}
  \REVIEW{Semiconductor Science and Technology} {27}{2012}{124003}

 \bibitem{Stanescu13}
  \Name{T. D. Stanescu and S. Tewari}
  \REVIEW{Journal of Physics: Condensed Matter} {25}{2013}{233201}

\bibitem{Elliott15}
  \Name{S. R. Elliott and M. Franz}
  \REVIEW{Rev. Mod. Phys.} {87}{2015}{137}
  
  \bibitem{Amet16}
  \Name{F. Amet, C. T. Ke, I. V. Borzenets, J. Wang, K. Watan-
abe, T. Taniguchi, R. S. Deacon, M. Yamamoto, Y. Bomze, S. Tarucha, et al.}
  \REVIEW{Science} {352}{2016}{966}
  
  \bibitem{Shalom16}
  \Name{M. Ben Shalom, M. J. Zhu, V. I. Falko, A. Mishchenko,A. V. Kretinin, K. S. Novoselov, C. R. Woods, K. Watanabe, T. Taniguchi, A. K. Geim, et al.}
  \REVIEW{Nat. Phys.} {12}{2016}{318}

\bibitem{Alicea16}
  \Name{J. Alicea and P. Fendley,}
  \REVIEW{ Annual Review of Condensed
Matter Physics} {7}{2016}{119}

\bibitem{Lee17}
  \Name{G.-H. Lee, K.-F. Huang, D. K. Efetov, D. S. Wei, S. Hart,
T. Taniguchi, K. Watanabe, A. Yacoby, and P. Kim}
  \REVIEW{Nat.Phys.} {13}{2017}{693}

\bibitem{Sahu18}
  \Name{M. R. Sahu, X. Liu, A. K. Paul, S. Das, P. Raychaudhuri, J. K. Jain, and A. Das}
  \REVIEW{ Phys. Rev. Lett.} {121}{2018}{086809}
  
\bibitem{Jeon19}
  \Name{G. S. Jeon, J. K. Jain, and C.-X. Liu,}
  \REVIEW{ Phys. Rev. B} {99}{2019}{094509}  
  
\bibitem{Chaudhary20}
  \Name{G. Chaudhary and A. H. MacDonald,}
  \REVIEW{ Phys. Rev. B} {101}{2020}{024516} 
  
  \bibitem{Geim92}
  \Name{A. K. Geim, S. J. Bending, and I. V. Grigorieva}
  \REVIEW{ Phys.Rev. Lett.} {69}{1992}{2252} 
  
  \bibitem{Wang13}
  \Name{L. Wang, I. Meric, P. Y. Huang, Q. Gao, Y. Gao, H. Tran,
T. Taniguchi, K. Watanabe, L. M. Campos, D. A.Muller, et al.}
  \REVIEW{ Science} {342}{2013}{614}
 
\bibitem{Sahu16}
  \Name{M. R. Sahu, P. Raychaudhuri, and A. Das,}
  \REVIEW{ Phys. Rev. B} {94}{2016}{235451} 

\bibitem{SM-Schirmer20}
  \Name{See Supplemental Material, which presents device char-
acterization, measurement methods, relevance of other
possible effects and the role of mesoscopic conductance
fluctuations. It also discusses the energy spectrum of the
theoretical model, the method by which the phases in
the MUC are determined, how the Hall conductivity is
computed, the Chern numbers of the Landau bands, and
more details about the density dependence of $\alpha'$}

\bibitem{Thouless82}
  \Name{D. J. Thouless, M. Kohmoto, M. P. Nightingale, and M. den Nijs}
  \REVIEW{ Phys. Rev. Lett.} {49}{1982}{405} 

\bibitem{Fukui05}
  \Name{T. Fukui, Y. Hatsugai, and H. Suzuki}
  \REVIEW{ Journal of the Physical Society of Japan} {74}{2005}{1674}

\bibitem{Nielsen95}
  \Name{M. Nielsen and P. Hedegard}
  \REVIEW{ Phys. Rev. B} {51}{1995}{7679}
  
  \bibitem{Galvis18}
  \Name{J. Galvis, E. Herrera, C. Berthod, S. Vieira, I. Guillamon, and H. Suderow,}
  \REVIEW{Communications Physics} {1}{2018}{1}
  
  \bibitem{Berdyugin19}
  \Name{A. I. Berdyugin, S. G. Xu, F. M. D. Pellegrino, R. Kr-
ishna Kumar, A. Principi, I. Torre, M. Ben Shalom, T. Taniguchi, K. Watanabe, I. V. Grigorieva, et al.,}
  \REVIEW{Science} {364}{2019}{162}
  
  \bibitem{Beenakker89}
  \Name{C. W. J. Beenakker and H. van Houten}
  \REVIEW{ Phys. Rev. Lett.} {63}{1989}{1857}

\end{thebibliography}
\end{document}


\title{Supplementary Materials for ``Hall Effect for Dirac Electrons in Graphene Exposed to an Abrikosov Flux Lattice"}

\author{Jonathan Schirmer$^1$$^*$, Ravi Kumar$^2$$^*$, Vivas Bagwe$^3$, Pratap Raychaudhuri$^3$, Takashi Taniguchi$^4$, Kenji Watanabe$^4$, C. -X. Liu$^1$, Anindya Das$^2$, J. K. Jain$^1$}
\affiliation{$^1$Department of Physics, 104 Davey Lab, The Pennsylvania State University, University Park, Pennsylvania 16802}
\affiliation{$^2$Department of Physics, Indian Institute of Science, Bangalore 560012, India}
\affiliation{$^3$Tata Institute of Fundamental Research, Homi Bhabha Road, Colaba, Mumbai 400 005, India}
\affiliation{$^4$National Institute for Materials Science, Namiki 1-1, Ibaraki 305-0044, Japan}



\maketitle

\setcounter{figure}{0}
\setcounter{table}{0}
\setcounter{equation}{0}
\renewcommand\thefigure{S\arabic{figure}}
\renewcommand\thetable{S\arabic{table}}
\renewcommand\theequation{S\arabic{equation}}

\section{Theory}
\subsection{Gauge choice}

\begin{figure*}
\includegraphics[width=0.7\textwidth]{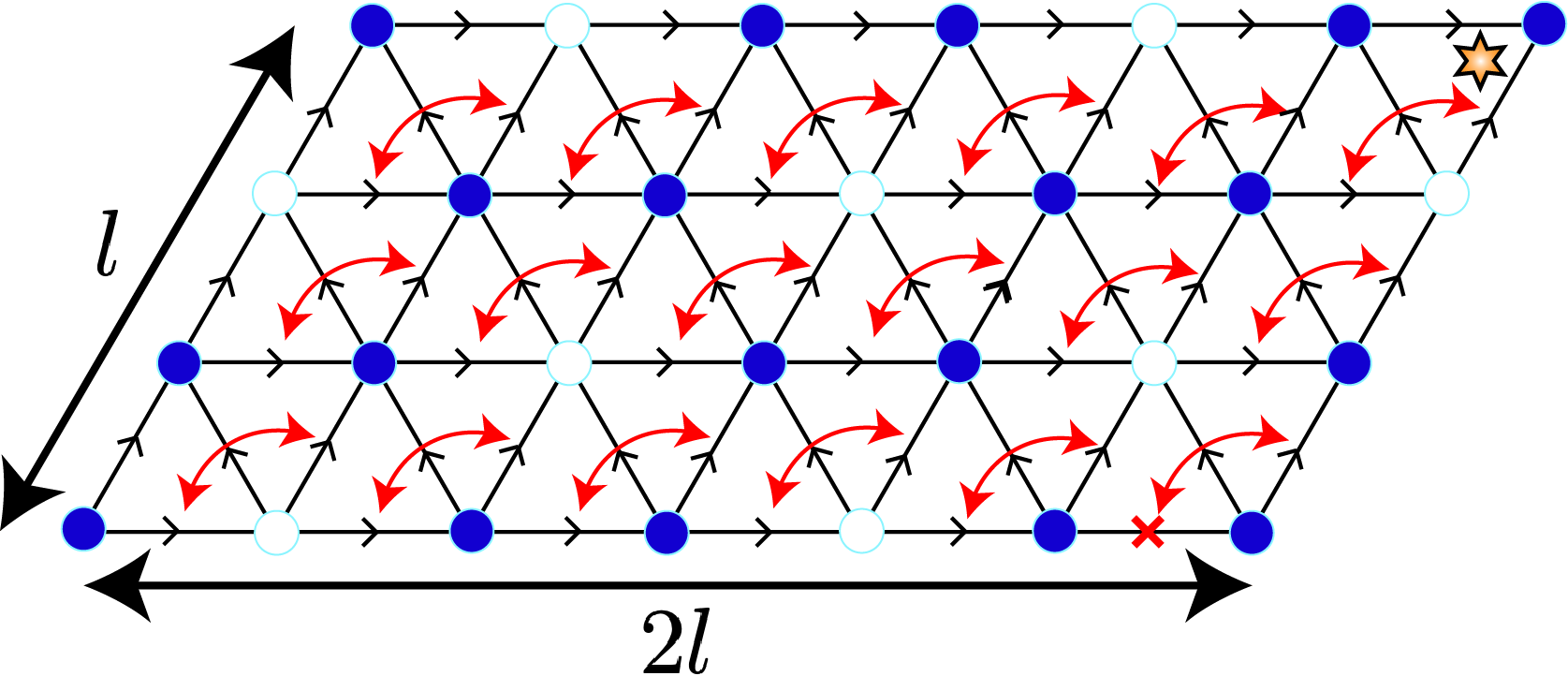}
\caption{The MUC for the parent triangular lattice. The white lattice sites are removed from the triangular lattice to produce a honeycomb lattice (Fig. \ref{phases_honeycomb}). To produce a periodic MUC for honeycomb lattice, $l$ must be divisible by 3 (shown here is $l=3$). Gauge choice: The phases connected by red arrows are set equal. The phase at the bottom right (at the {\color{red}$\mathbf{\times}$}) is set to zero. A point flux of $-\Phi_0$ is placed at the location of the orange star. }  \label{fig1}
\end{figure*}

\begin{figure*}
\includegraphics[width=0.7\textwidth]{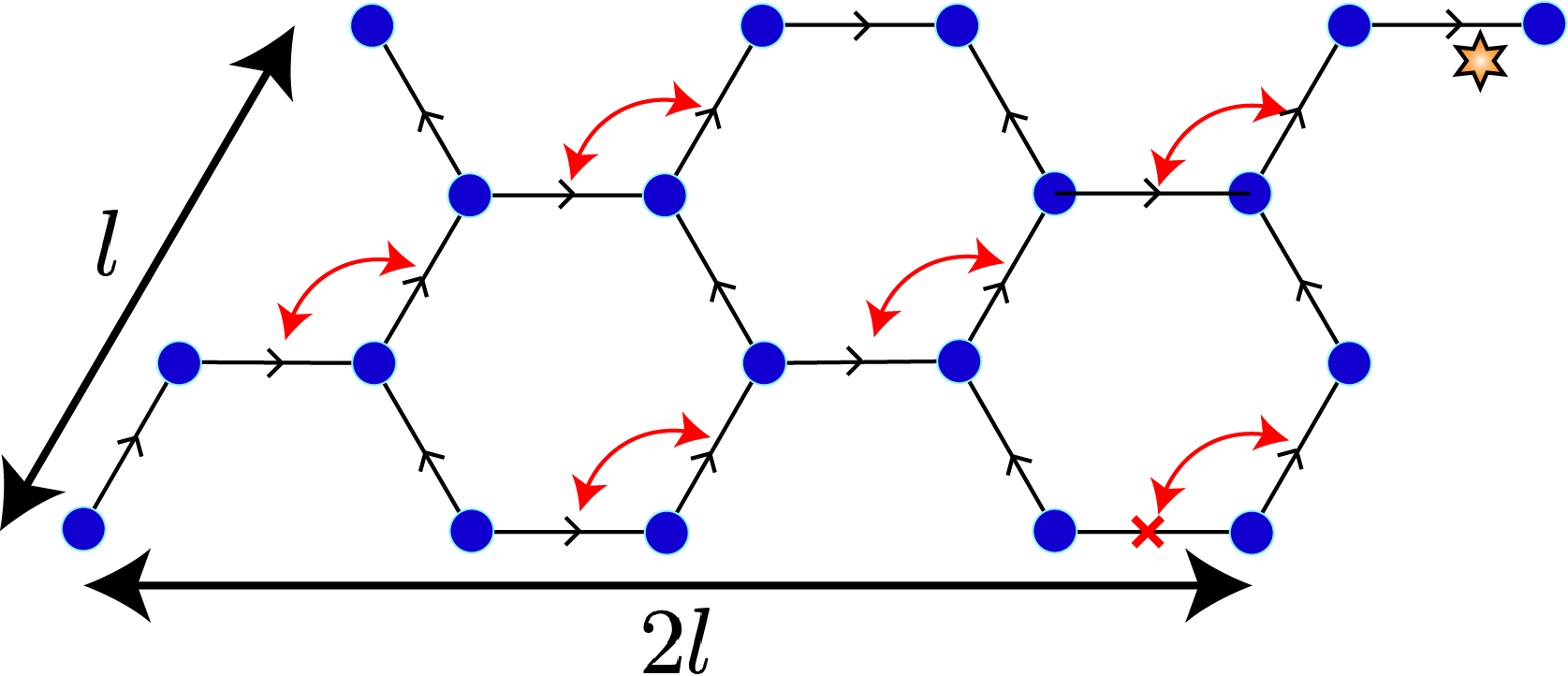}
\caption{The honeycomb lattice MUC produced after the appropriate lattice sites are removed from the triangular lattice MUC. The phases connected by red arrows are equal. The phase at the bottom right (at the {\color{red}$\mathbf{\times}$}) is zero. A point flux of $-\Phi_0$ is placed at the location of the orange star. }  \label{phases_honeycomb}
\end{figure*}

Here we show how we determine the phases of the hopping parameters within a magnetic unit cell (MUC). We will discuss how it is done for of a triangular lattice; the honeycomb lattice is obtained by removing some of the sites (see Fig. \ref{fig1} and Fig. \ref{phases_honeycomb}).
 We will assume that the MUC is of size $2l\times l$, where $l$ is an integer quoted in units of $a$, which is the lattice constant of the discretized lattice (the distance between two neighboring sites in Fig.~\ref{fig1}). Furthermore, in order that a honeycomb MUC may be generated from the parent triangular MUC by removal of sites, we require that $l$ be divisible by 3.

The phases associated with each nearest-neighbor hopping within the magnetic unit cell are determined by the flux penetrating through each triangular plaquette, together with a gauge-fixing condition. In particular, for a given triangle $j$, we have
\begin{equation} \label{eq6}
\sum_{m,n \in \triangle_j}\theta_{m,n}=2\pi \frac{\Phi_j}{\Phi_0}
\end{equation}
where $\theta_{m,n }$ is the phase associated with hopping from site $n$ to $m$, $\Phi_j$ is the flux through triangle $j$ (denoted as $\triangle_j$), $\Phi_0=h/e$, and the sum is taken in a counter-clockwise sense. Taking sum of the equations for each $\vartriangle_j$ gives
\begin{equation} \label{bound}
\sum_{m,n \in {\partial {\rm MUC}}}\theta_{m,n}=2\pi \frac{\Phi_{\rm tot}}{\Phi_0}
\end{equation}
where the sum is taken along the boundary of the MUC (denoted $\partial {\rm MUC}$). We wish to make a periodic choice for the phases, but that would require that the left hand side of the above equation, and hence the net flux through the MUC, vanish. In our case, the net flux through the MUC is one flux quantum. To allow for periodic choice of phases, we apply a flux of $-\Phi_0$ through one triangle so as to make the total applied flux equal to zero (see Fig. \ref{fig1}). The additional $-\Phi_0$ flux has no physical effect.

Within a MUC, there are $6l^2$ independent phases and $4l^2-1$ equations of the type Eq. \ref{eq6} determining them. Therefore Eq. \ref{eq6} does not uniquely determine the phases $\theta_{m,n}$; an additional $2l^2+1$ constraints are required to fix the phases, which are equivalent to a choice of gauge. The gauge chosen in our model is the following: for any given lattice site, we set the phases associated with hopping into it from the left
and hopping out of it in the direction $\frac{1}{2}\hat{\vec{x}}+\frac{\sqrt{3}}{2} \hat{\vec{y}}$ equal (these phases are connected by red arrows in Fig. \ref{fig1}). Further, the phase for hopping
marked by a red {\color{red}$\mathbf{\times}$} in Fig. \ref{fig1} is set to zero. This gives $2l^2+1$ additional constraints, as required.

\subsection{Spectrum}

The Hamiltonian of our model consists of nearest-neighbor hopping matrix elements with phases determined via the method described in the previous section. Because of periodicity, we are able to write the Hamiltonian in momentum space, with the MUC as the unit cell:
\begin{equation}
\hat{H}=\sum_{\vec{k}}\sum_{m,n \in {\rm MUC}} {c^\dagger_{\vec{k},m} h_{mn}(\vec{k}) c_{\vec{k},n}}
\end{equation}
where $c^\dagger_{\vec{k},m}$ ($c_{\vec{k},m}$) creates (annihilates) an electron at momentum $\vec{k}$ at site $m$ within the MUC. Diagonalization of the matrix $h_{mn}(\vec{k})$
yields the energy levels at a fixed $\vec{k}$.

\begin{figure*}
\includegraphics[width=.45\textwidth]{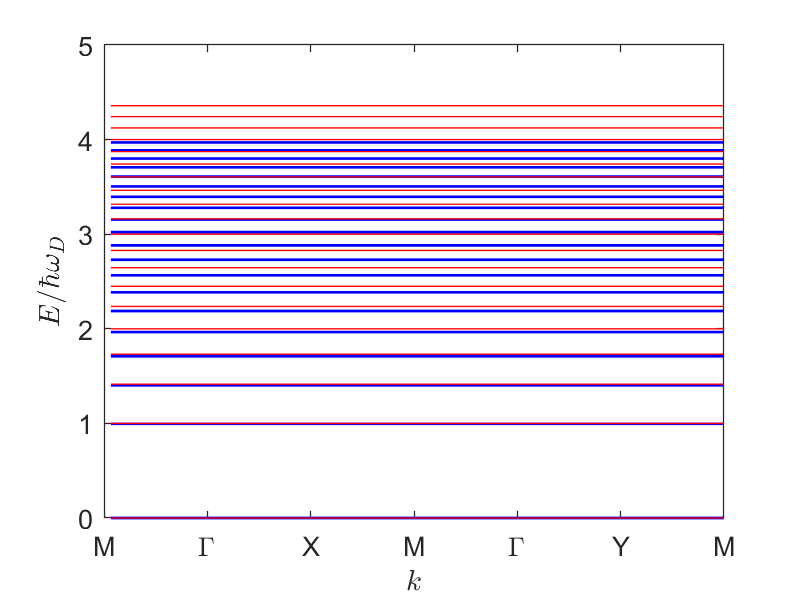}
\includegraphics[width=.45\textwidth]{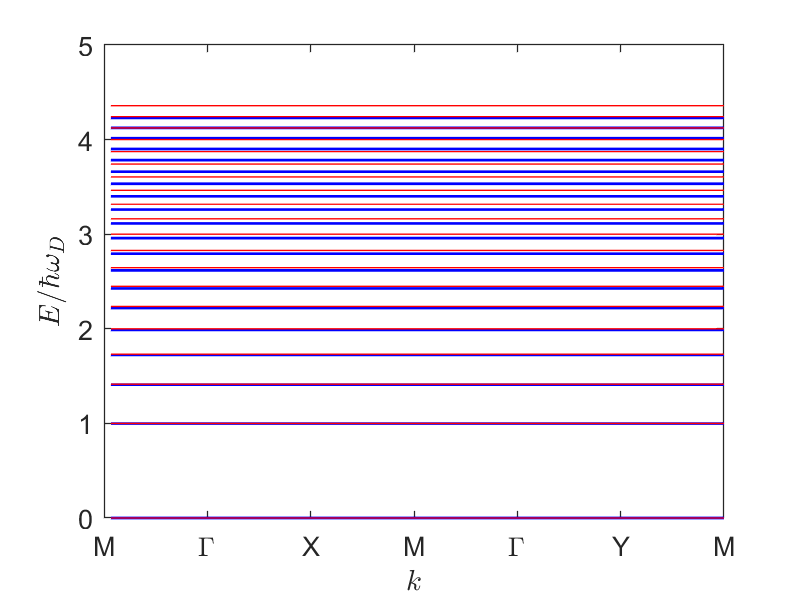}
\caption{The spectrum of the honeycomb lattice model in a uniform magnetic field for a MUC of dimensions $36\times18$, i.e. $l=18$ (top) and $60\times30$, i.e. $l=30$ (bottom). The energy scale $\hbar\omega_D$ denotes the gap between the $n=1$ and $n=0$ levels in uniform magnetic field and in the continuum limit.  The red lines denote the energy levels in a uniform magnetic field in the continuum limit (the $n^{\rm th}$ level lies at precisely $E={\rm sgn}(n)\sqrt{|n|}\hbar\omega_D$, as appropriate for a linear dispersion relation). Note that each level consists of a pair of bands. In both panels, levels from $n=0$ to $n=20$ are shown. The deviation from the continuum behavior becomes less pronounced, as expected, as $l$ is increased.} \label{l_compare}
\end{figure*}

\begin{figure*}
\includegraphics[width=.45\textwidth]{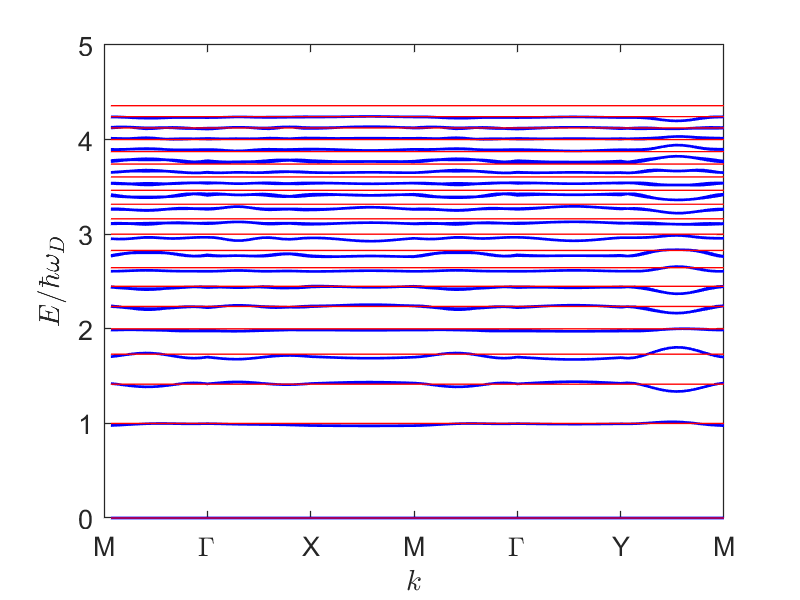}
\includegraphics[width=.45\textwidth]{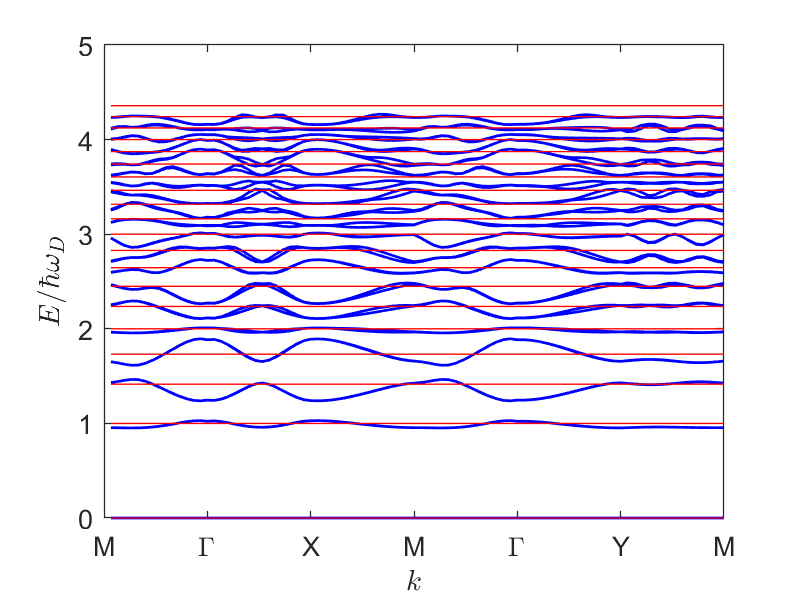}
\caption{The spectrum of the honeycomb lattice model for $d/l=0.2$ (left), and $d/l=0.02$ (right) for a calculation with MUC of dimensions $60\times30$, i.e. $l=30$. The energy scale $\hbar\omega_D$ denotes the gap between the $n=1$ and $n=0$ levels in uniform magnetic field and in the continuum limit. The red lines mark the energy levels in a uniform magnetic field in the continuum limit (the $n^{\rm th}$ level lies at precisely $E={\rm sgn}(n)\sqrt{|n|}\hbar\omega_D$, as appropriate for a linear dispersion relation). In both panels, levels from $n=0$ to $n=20$ are shown. }  \label{fig2}
\end{figure*}

In our calculations, which are intended to model graphene, we focus on positive energies near $E=0$ in the spectrum of the honeycomb lattice. Concerning the spectrum of this model, several things are of note. In uniform magnetic field, the spectrum near
the $E=0$ consists of doubly degenerate (due to valley degeneracy) and flat Landau bands. The LLs of Dirac electrons with perfectly linear dispersion occur at energies proportional to ${\rm sgn}(n)\sqrt{|n|}$, where $n \in \mathbb{Z}$ is the LL index. For the first few low-energy Landau bands in our calculation, the energies closely coincide with the $\sqrt{n}$ scaling, as seen in Fig.~\ref{l_compare}; the ideal LLs of the Dirac particles are indicated
red lines in Fig.~\ref{l_compare}. At somewhat higher energies, there is some deviation from the $\sqrt{n}$ behavior, indicating a slight deviation from a linear dispersion relation at these energies. However, we note that as we increase $l$ (and thus move toward the continuum limit), the deviation is pushed to higher and higher LLs.
As the magnetic field is made non-uniform, the Landau bands acquire a finite width but, somewhat surprisingly, maintain their band center energies near their uniform field values. Additionally, as the magnetic field is made highly non-uniform, the bands at $n=0$ remain flat to a remarkable degree. This is seen most sharply in the rightmost panel of Fig. \ref{fig2} where $d/l=0.02$, i.e. the spacing between the Abrikosov vortices is 50 times their size. Therefore, we anticipate that the QH effect due to the low filling Landau levels ought to be observable even when the size of the Abrikosov fluxes is much smaller than their separation. This theoretical prediction requires future experimental examination.

\subsection{Computation of Hall conductivity}

\begin{figure*}
\includegraphics[width=.65\linewidth]{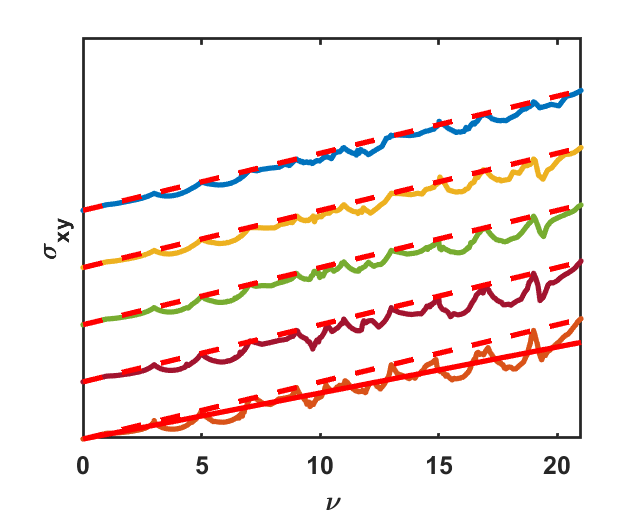}
\caption{The Hall conductivity for $d/l$ values not included in the main article. The various lines correspond (from lowest to highest) to $d/l=$ 0.02, 0.06, 0.1, 0.14, 0.18. Linear fitting for conductivity is shown only for the lowest curve (solid red line).}\label{sigma}
\end{figure*}

In order to numerically calculate the Hall conductivity, we discretize the Brillouin zone into a grid. The Brillouin zone can be broken into $R^2$ points by defining the grid $\vec{k}_{\vec{\alpha}}=k_{\alpha_1} \hat{\vec{k}}_1+k_{\alpha_2} \hat{\vec{k}}_2$, where
\begin{align}
k_{\alpha_1}&=\frac{2 \pi }{\sqrt{3}lR}\alpha_1 - \frac{\pi}{\sqrt{3}l} \quad (\alpha_1=0,\dots,R-1) \\
k_{\alpha_2}&=\frac{4 \pi}{\sqrt{3}lR}\alpha_2 - \frac{2\pi}{\sqrt{3}l} \quad (\alpha_2=0,\dots,R-1)
\end{align}
Here we have $\vec{\alpha}=(\alpha_1,\alpha_2)$, and $\hat{\vec{k}}_1=\frac{\sqrt{3}}{2}\hat{\vec{x}}-\frac{1}{2}\hat{\vec{y}}$, $\hat{\vec{k}}_2=\hat{\vec{y}}$ are magnetic reciprocal lattice unit vectors. For our calculations, we chose $R=30$, where we find the results to be well converged. We also define a set of vectors $\vec{e}_\lambda$, $\lambda=1,2$:
\begin{align}
\vec{e_1}&=\frac{2 \pi }{\sqrt{3}lR} \hat{\vec{k}}_1 \\
\vec{e_2}&=\frac{4 \pi }{\sqrt{3}lR} \hat{\vec{k}}_2
\end{align}
which give the spacing between points in the Brillouin zone.
We then define link matrices for a chosen set of $K$ bands labeled by $i,j=1,\ldots,K$ at each point on the grid:
\begin{equation}
M^{ij}_\lambda(\vec{k_{\alpha}})=\langle u^i (\vec{k_{\alpha}})| u^j (\vec{k_{\alpha}}+\vec{e}_{\lambda}) \rangle \label{eq19}
\end{equation}
 where $|u^i(\vec{k})\rangle$ is the Bloch state within band $i$ and $\lambda=1,2$. In terms of the link variables defined as
\begin{equation}
U_\lambda(\vec{k_\alpha})=\frac{\det M_\lambda(\vec{k_\alpha})}{|\det M_\lambda(\vec{k_\alpha})|}
\end{equation}
 the discrete Berry curvature at each point on the grid is given by
\begin{multline}
\tilde{F}(\vec{k_{\alpha}})= \ln\Big( U_1(\vec{k_{\alpha}})  U_2(\vec{k_{\alpha}}+\vec{e_1})  \\ U_1^{-1}(\vec{k_{\alpha}}+\vec{e_2}) U_2^{-1}(\vec{k_{\alpha}}) \Big)
\end{multline}
The Hall conductivity is then given by
\begin{equation} \label{eq22}
\sigma_{xy}=\frac{1}{2\pi i}\sum_{\vec{\alpha}}\tilde{F}(\vec{k_{\alpha}})
\end{equation}
where the sum is over occupied states in the Brillouin zone, i.e. only those values of $\alpha$ are chosen for which
$E(\vec{k_{\alpha}})<E_{\rm F}$, where $E_{\rm F}$ is the Fermi energy. For all bands which fully lie below $E_{\rm F}$, the contribution to $\sigma_{xy}$ will be an integer-valued Chern number $C$. Partially occupied bands (i.e. bands which cross the Fermi level) will have a contribution to $\sigma_{xy}$ which is, in general, not an integer. Fig. \ref{sigma} shows $\sigma_{xy}$ as a function of filling factor $\nu$ for several $d/l$ values not shown in Fig. 2a of the main article.

\subsection{Chern numbers and physically relevant bands}

\begin{figure*}
\centering

\includegraphics[width=.45\linewidth]{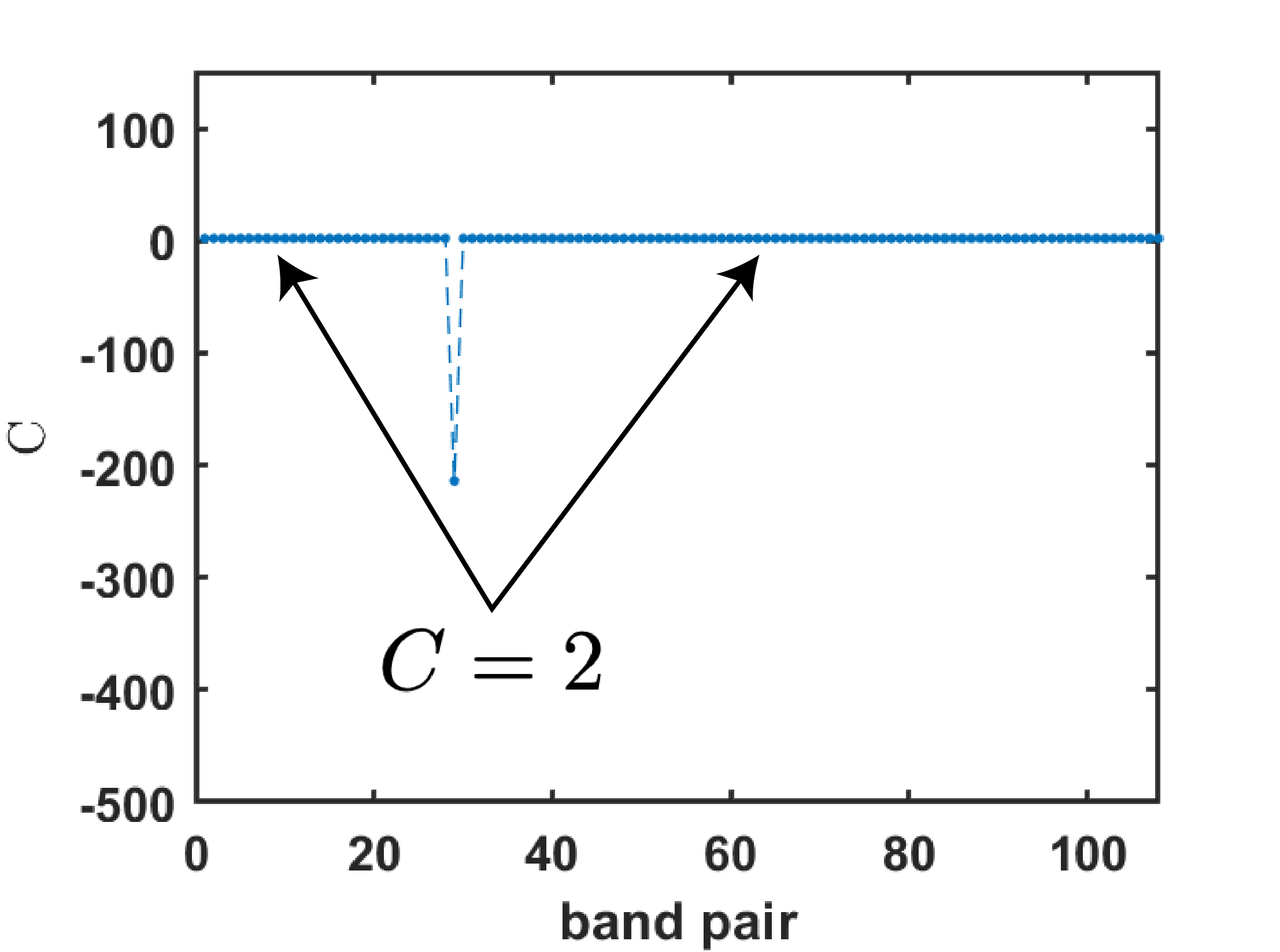}
\includegraphics[width=.45\linewidth]{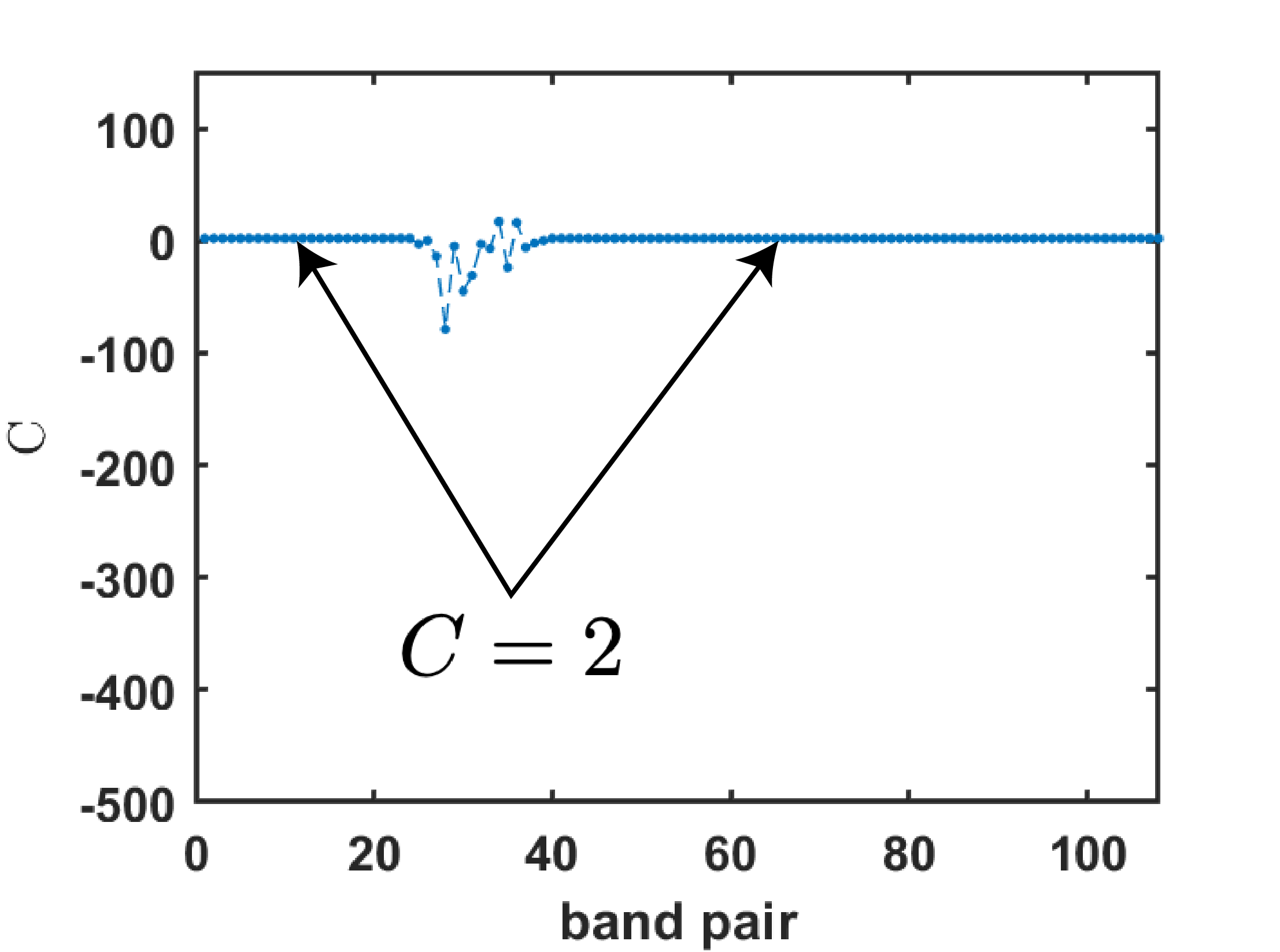}
\includegraphics[width=.45\linewidth]{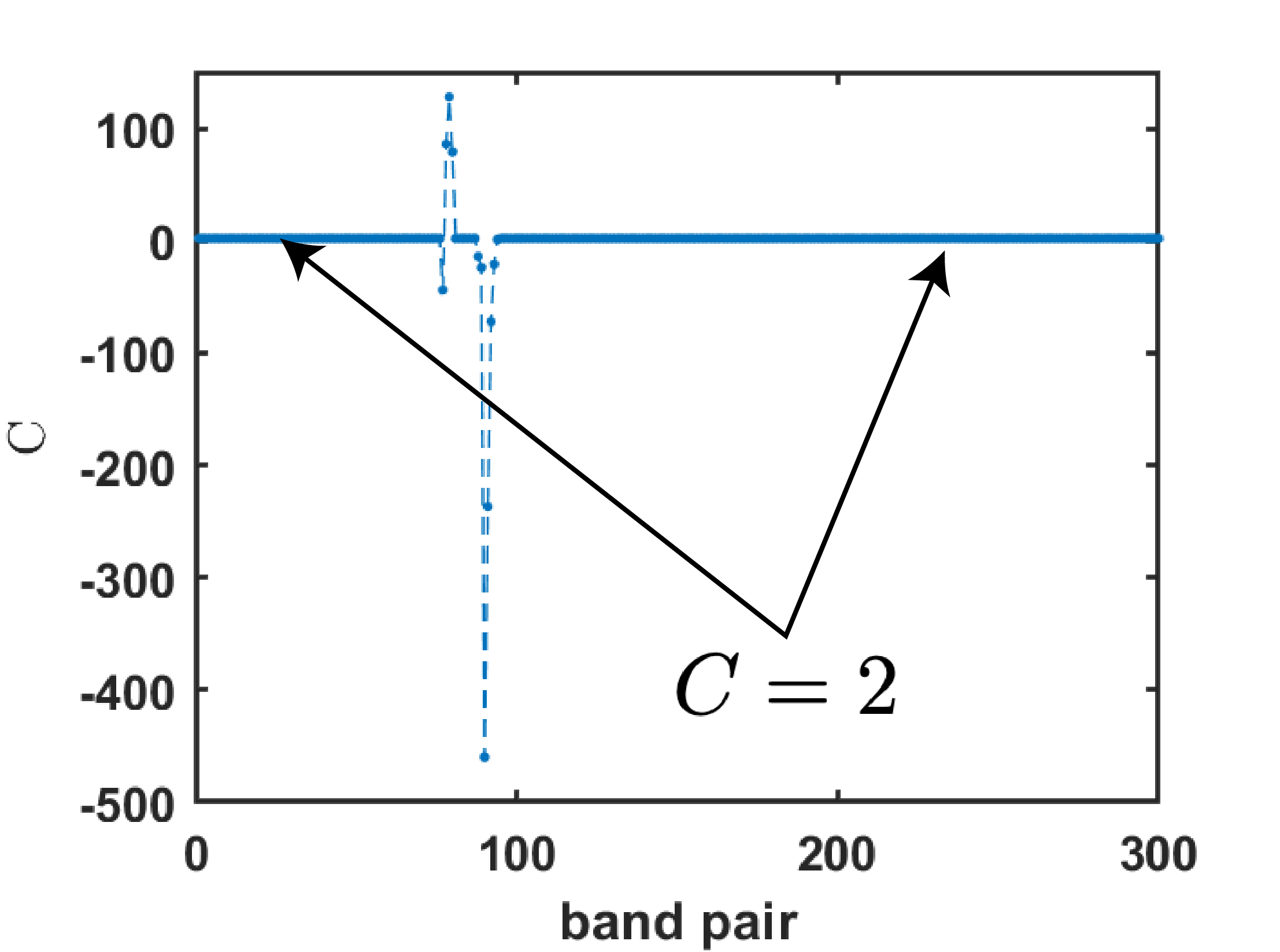}
\includegraphics[width=.45\linewidth]{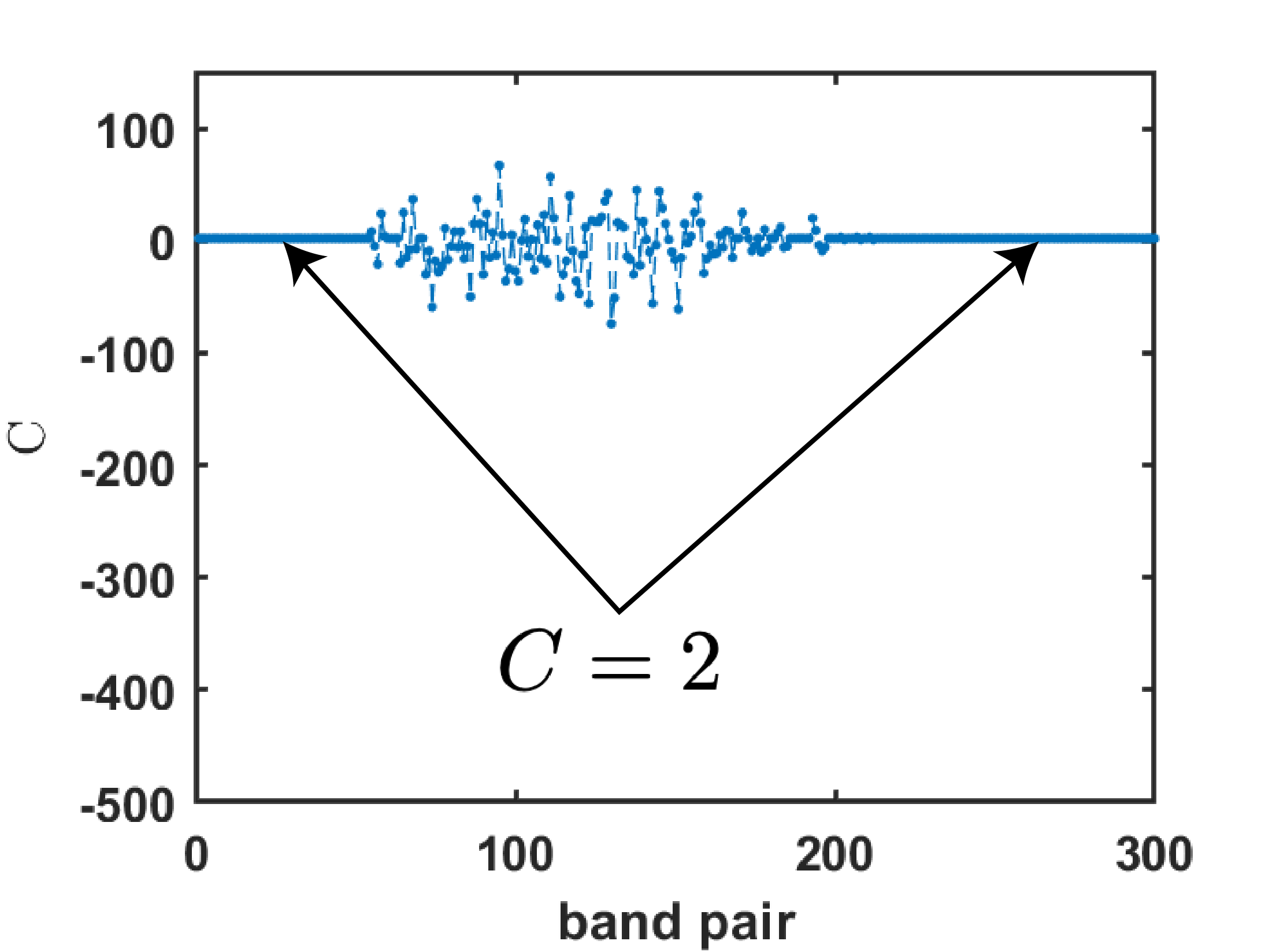}

\caption{The left column shows the Chern numbers of pairs of bands for a uniform magnetic field with MUC of dimensions $36\times18$, i.e. $l=18$ (top) and $60\times30$, i.e. $l=30$ (bottom). Note that band pair $n$ occurs at filling factor $\nu=2n+1$ because half of the pair at $n=0$ belongs to the valence band. The right column shows the Chern numbers of pairs of bands when $d/l=0.02$ with MUC of dimensions $36\times18$, i.e. $l=18$ (top) and $60\times30$, i.e. $l=30$ (bottom). The band pairs at low and high energies have $C=2$, but there is a range of energies where bands with very high values of $|C|$ occur; these bands are pushed to higher and higher fillings as the continuum limit is approached, and are thus not relevant to the physics under investigation.} \label{CN}
\end{figure*}

It is important to ensure that our results are reflective of the continuum limit. For that purpose, we ask how the Chern numbers of various bands evolve as we make the lattice finer.
Fig. \ref{CN} shows the Chern numbers ($C$), for $2l\times l$ lattices with $l=18$ and $l=30$, for both uniform and non-uniform magnetic fields. In each case, we give $C$ for pairs of degenerate or quasi-degenerate bands occurring at $n=0$ up to the maximum filling. Several features are worth noting.  In both uniform and non-uniform magnetic fields, the Chern number of each degenerate pair of bands remains 2 (i.e. each band has on average unit Chern number) for $|n| \leq 27$ for $l=18$ and $|n|\leq 76$ for $l=30$. Beyond that, a few bands in a relatively narrow energy range carry Chern numbers with very large magnitude. At yet higher energies, the bands resume the  normal behavior with $C=2$. The bands with anomalous Chern numbers occur when the dispersion relation for electrons in a honeycomb lattice crosses from linear, which occurs near $E=0$, to parabolic, which occurs near the maximum of the conduction band.
As the continuum limit is approached with increasing $l$, these anomalous bands rise to ever higher fillings, leaving behind bands which, together with their valley degenerate partner, carry $C=2$. For our calculations,
we are careful to consider fillings low enough so as to avoid the anomalous bands with high $C$ for all values of $l$ as we approached the continuum limit.

\subsection{Dependence of $\alpha'$ on density}

\begin{figure*}
\includegraphics[width=.65\linewidth]{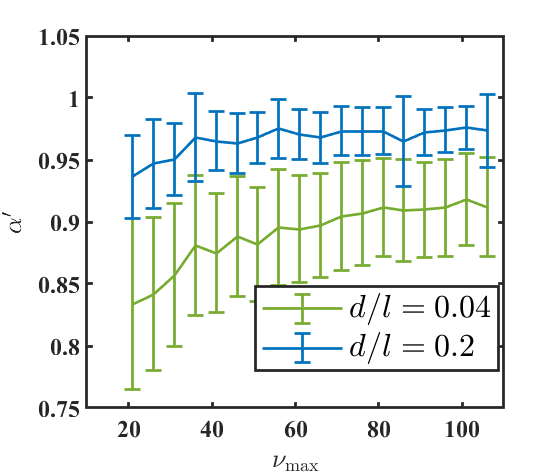}
\caption{The quantity $\alpha'$ as a function of the range of fillings (from $\nu=0$ to $\nu=\nu_{\rm max}$) over which $\sigma_{xy}$ is fitted. A $2l\times l$ unit cell with $l=30$ is considered. The density ranges are $6\times 10^{10}$ cm$^{-2}$ to  $3\times 10^{11}$ cm$^{-2}$ for the blue curve and $2.4\times 10^{9}$ cm$^{-2}$ to  $1.2\times 10^{10}$ cm$^{-2}$ for the green curve. We note that these results do not reflect the continuum limits.}  \label{nu_max}
\end{figure*}

As pointed out in the main article, our experiments found that $\alpha'$ increases with the density of charge carriers. In order to extrapolate $\alpha'$ to the continuum limit in our theoretical calculation, we have fitted $\sigma_{xy}$ over a range of filling factors, from $\nu=0$ to $\nu=\nu_{\rm max}$, which stays sufficiently far from the bands with high $|C|$ for all values of $l$. We considered $l=$ 18, 21, 24, 27, 30 and chose $\nu_{\rm max}=21$ for the extrapolation to the continuum limit.
Accordingly, our results in the continuum limit are for low densities (in the range 6$\times$10$^{8}$ cm$^{-2}$ $-$ 6$\times$10$^{10}$ cm$^{-2}$). However, we find for finite $l$, that $\alpha'$ displays dependence on $\nu_{\rm max}$ and, in fact, increases with $\nu_{\rm max}$ (see Fig. \ref{nu_max}). Therefore, this suggests that $\alpha'$ should increase with the density of charge carriers, as seen in our experiment. It should be noted, given the results shown in Fig. \ref{nu_max}, that is not possible to determine the value of $\alpha'$ in the high density limit since that would require a calculation with $\nu_{\rm max}=5\times 10^{5}$  (given the experimental parameters), which is outside our computational capability.  Nevertheless, the calculations do not rule out that $\alpha'\rightarrow 1$ in the high density limit, as our experiment suggests.

\section{Experiment}
\subsection{Device characterization}
\begin{figure*}
 \includegraphics[width=\linewidth]{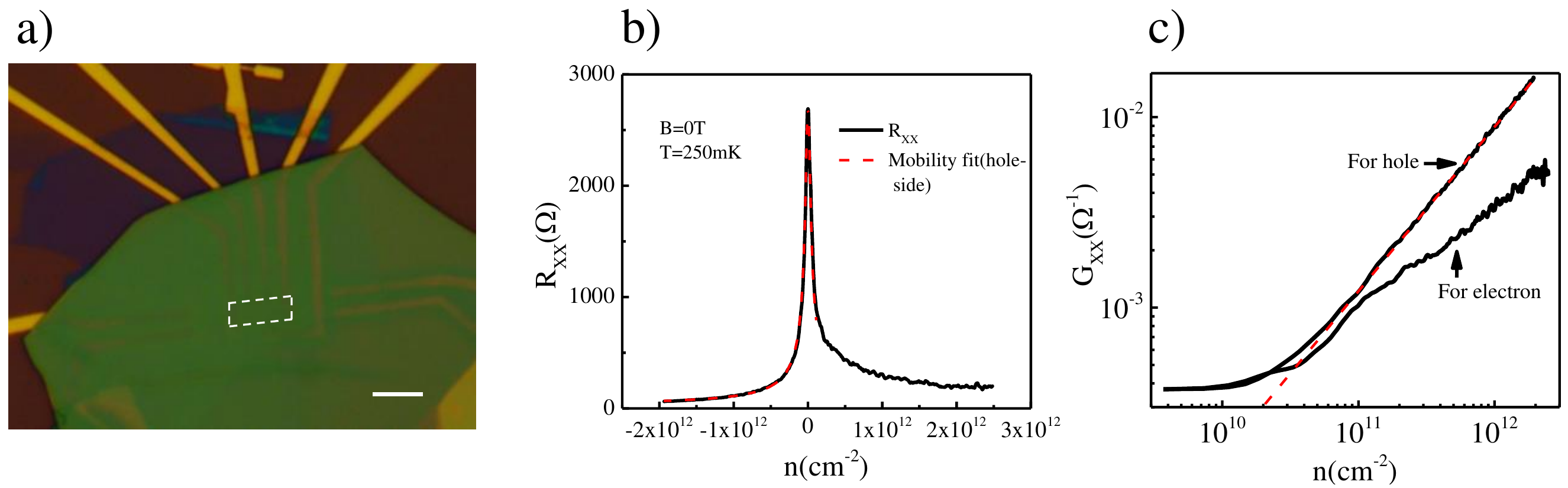}
  \caption{a) Optical image of the measured graphene-hBN-NbSe$_{2}$ heterostructure. The white dashed line represents the boundary of graphene channel. The scale bar is 3$\mu$m. b) Four probe gate response of the graphene channel at $T=250$ mK and $B=0$ T. The dashed red line represents the fitting for the hole side using Eq. \ref{eq1}. c) The conductance versus carrier density on a log-log scale. The red dashed line is the linear fit used to obtain the charge inhomogeneity.}
\label{expfig1}
\end{figure*}

Fig.~\ref{expfig1}a shows the optical image of our graphene-hBN-NbSe$_{2}$ heterostructure. The white dashed line represents the boundary of the graphene channel. The gold colored lines are the contacts for graphene and NbSe$_{2}$. The green flake on the top is $\sim$ 40nm thick NbSe$_2$, which covers the entire graphene channel. The purple thin flake corresponds to $\sim$ 10 nm thick hBN, which electrically isolates the graphene and its contacts from the NbSe$_2$ flake.

Fig.~\ref{expfig1}b shows the four-probe resistance of the graphene channel as a function of back-gate voltage. The measured four-probe resistance is fitted with the following equation:
\begin{equation}\label{eq1}
{\rm }R=\frac{L}{We{\mu}{(n_{0}^2+n^2)}^{1/2}}
\end{equation}
where $W$ is width of the sample, $L$ is length between the longitudinal voltage probes, $\mu$ is mobility, $e$ is electron charge, $n$ is the charge carrier concentration induced by the back gate voltage, and $n_{0}$ is a measure of charge inhomogeneity due to electron-hole charge puddles. The carrier concentration is given by $n$=${\rm}C_{BG}(V_{BG}-V_{D})$, where $C_{BG}$ is the back gate capacitance of $300$~nm thick SiO$_{2}$, $V_{BG}$ is the voltage applied to the gate, and $V_{D}$ is the back-gate voltage corresponding to the Dirac point. $W$ and $L$ are 2.1 $\mu$m and 2 $\mu$m, respectively. The fitting gives mobility $\sim$ 68,000 cm$^2$/V/s and charge inhomogeneity $n_{0}$ $\sim$ 3$\times$10$^{10}$ cm$^{-2}$.

In Fig.~\ref{expfig1}c, we have plotted the conductance $G$ versus carrier density $n$ on a log-log scale. Conductance increases linearly with $n$ after $n$ sufficiently exceeds $n_0$, as seen by the red dashed line for hole side in Fig.~\ref{expfig1}c. Note that in our experiment the working range of carrier concentration is $\sim$ 1$\times$10$^{11}$ cm$^{-2}$ to 1.7$\times$10$^{12}$ cm$^{-2}$, where the conductance is linear in $n$.

\subsection{Measurement of Hall voltage }

\begin{figure*}
 \includegraphics[width=\linewidth]{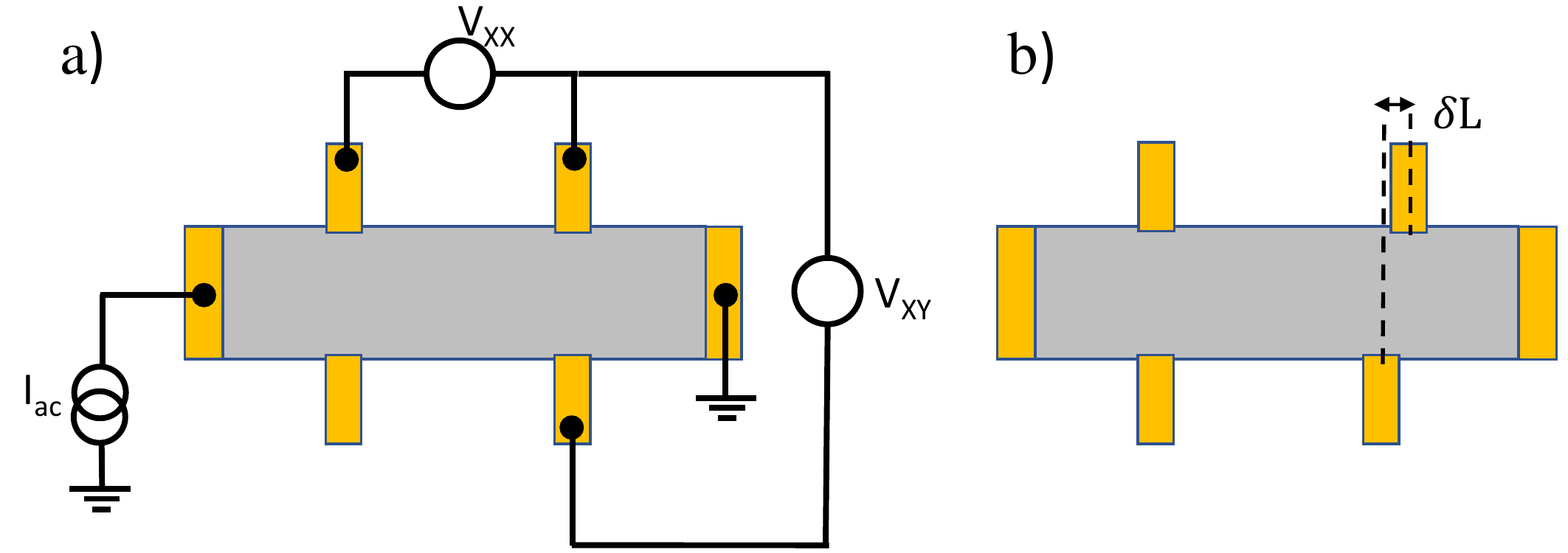}
  \caption{a) Measurement schematic. b) Possible misalignment in transverse probes due to lithographic error (labeled $\delta L$).}
  \label{expfig2}
\end{figure*}

The graphene channel was current biased, and $V_{xx}$ and $V_{xy}$ were simultaneously measured using lock-in amplifiers as shown in Fig.~\ref{expfig2}a. A lithographical uncertainty on the order of $50-100$ nm is unavoidable, which can give a longitudinal shift in the transverse probes as shown in Fig.~\ref{expfig2}b. Hence, the measured $R_{xy}$ can have a small longitudinal component. We eliminate it in the standard fashion, by considering an appropriate average of $R_{xy}$ at $\pm B$:
\begin{equation}
R_{xy}(B)={R_{xy}(+B)-R_{xy}(-B)\over 2}. \label{eq:Rxy}
\end{equation}
 This contains only the transverse component, because
the measured $R_{xy}$ for $\pm B$ magnetic field is given by
$R_{xy}(\pm B)=\frac{\pm B}{ne}+\frac{\rho (\delta L)}{W}$,
where $\delta L$ is the shift in the transverse probes and $\rho$ is the longitudinal resistivity.
$R_{xy}(B)$ defined in Eq.~\ref{eq:Rxy}, which will be referred to as the ``average" $R_{xy}$, is
shown in Fig.~\ref{expfig3} and Fig.~\ref{expfig4}.
The reduction of the Hall resistance  at low concentrations and low magnetic fields can be seen from the ratio of $R_{xy}$ below and above $T_{C}$ (called $\alpha$).

The elimination of the $R_{xx}$ component in the above analysis assumes that
$R_{xx}$ does not depend on the direction of magnetic field. In Fig.~\ref{expfig5}, we have replotted the average $R_{xy}$ at 20 mT and 40mT, along with the measured $R_{xx}$ for $\pm$ 20mT and $\pm$40mT, for temperatures above and below $T_{C}$. The inset shows the ratio $R_{xx}(+B)/R_{xx}(-B)$
as a function of the carrier concentration. The ratio is seen to be equal to unity within experimental accuracy.

\subsection{Conversion of Hall resistance to Hall conductivity}

The Hall component of the conductivity is given by
\begin{equation}
{\rm }\sigma_{xy}=\frac{-\rho_{xy}}{\rho_{xx}^2+\rho_{xy}^2}
\end{equation}
where $\rho_{ xx}=\frac{R_{xx} W}{L}$ and $\rho_{ xy}={\rm }R_{xy}=\frac{B}{ne}$.
At the magnetic fields (10-100 mT) used in our experiment, we have $R_{xx} > R_{xy}$, as can be seen from Fig.~\ref{expfig5}, implying $R_{xx}^2 \gg R_{xy}^2$. As a result, we have $\sigma_{xy} \propto R_{xy}$. 

\subsection{Relation between $d$, $l$ and $B$}

The quantity $d$ is the diameter of the Abrikosov vortex core, $l$ is the lattice constant of the Abrikosov flux lattice, and $B$ is the applied magnetic field. The lattice constant $l$ is given by
\begin{equation}
{\rm}l=1.075\times\Big({\frac{\Phi_0}{2B}}\Big)^{1/2}
\end{equation}
where $\Phi_{0}=\frac{h}{e}$. The typical value of $d \sim 30nm$\cite{galvis2018tilted}. 

\subsection{Other possible effects}

In this section we discuss hydrodynamic \cite{berdyugin2019measuring}, ballistic \cite{beenakker1989billiard} and proximity effects, and argue
that these are not relevant to the reduction of the Hall resistance observed in our experiments.

The hydrodynamic regime dominates when the electron-electron scattering length ($l_{e-e}$)
is the shortest length scale in the problem, so that $l_{e-e} \ll W,\,l,\, l_{e-ph}$, where $W$ is the sample size, $l(=v_{f}\tau$) is the mean free path due to disorder scattering, and $l_{e-ph}$ is the electron-phonon scattering length. It has been shown in the literature \cite{berdyugin2019measuring} that the hydrodynamic effects in graphene are observed only in extremely clean samples (mobility $\sim$ 1$\times$10$^{5}$cm$^{2}$V$^{-1}$s$^{-1}$ - 5$\times$10$^{5}$V$^{-1}$s$^{-1}$) and in intermediate temperature range (50-100K) \cite{berdyugin2019measuring,bandurin2016negative,crossno2016observation}. In contrast, in our experiment, the mobility of the device is in moderate range ($\sim$ 6.8$\times$10$^{4}$cm$^{2}$V$^{-1}$s$^{-1}$), and the working temperature is low ($250$mK to $15$K); the observed suppression in the measured Hall resistance occurs only at temperature $T<T_{C} (10K)$. Furthermore, correction to Hall resistance due to hydrodynamic effect is expected to vanish at a distance $\geq 1\mu$m away from the current injection contact \cite{berdyugin2019measuring}. In our device, the distance of the Hall probes were at $\sim 2\mu$m away from the current injection contact.
Based on these considerations we eliminate the possibility of hydrodynamic correction to the measured Hall resistance in our experiment.

Ballistic effects have been shown to suppress Hall resistance in mesoscopic sample in narrow channels \cite{beenakker1989billiard}, where the electron mean free path (i.e. $l$ =$v_{f}\tau , v_{f}=10^{6}ms^{-1}$) is larger than the dimensions of the device. However, for our device the mean free paths of the carriers are smaller than the sample dimensions (width $\sim 2.1\mu$m, channel length $\sim 6\mu$m) as shown in Table \ref{tab:table1}.  Thus, ballistic effects are unlikely to be relevant to the observed reduction in the Hall resistance data in our experiment as the magnetic field is made non-uniform.

\begin{table}[h!]
  \begin{center}
    \caption{The mean free path $l$, calculated for several carrier concentrations for two different temperatures of $250mK$ and $15K$. The mean free paths are small compared to the dimensions of our sample $\sim 2.1\mu$m $\times 6\mu$m.}
    \label{tab:table1}
    \begin{tabular}{|c|c|c|} 
\hline
      density [10$^{11}$cm$^{-2}$] & $l(\mu m)$ at 250mK & $l(\mu m)$ at15K \\
      \hline
      17 & 1.0 & 0.98\\
      10 & 0.78 & 0.75\\
      6 & 0.61 & 0.58\\
      3 & 0.45  & 0.41 \\
      1 &  0.31 & 0.24\\ \hline
    \end{tabular}
  \end{center}
\end{table}

Another effect that can affect transport is the proximity effect. We now argue that thickness of hBN ($\sim$ 10 nm) is large enough to suppress the proximity effect. For this purpose, we refer to the existing literatures, where the hBN has been used extensively as an insulating barrier or dielectric in mesoscopic systems \cite{dean2010boron}. hBN has a band gap of 5.2-5.9eV which makes it a perfect insulator. There are studies on the tunneling conductance of hBN as a tunnel barrier \cite{chandni2015evidence}. In Fig. 2b of reference \cite{chandni2015evidence}, tunneling conductance at zero bias has been shown to decrease exponentially with the number of layers, as expected quantum mechanically. From this study, one can estimate the tunneling conductance for a 10nm ($\sim$ 30 layers) thick hBN for our device to be $\sim$ 1$\times$10$^{-40}$S. Furthermore, in  Ref.~ \cite{dvir2018spectroscopy} the proximity effect of NbSe$_{2}$ via an insulating barrier (4-5 layers MoS$_2$) has been studied, where also one can see (Fig. 1b of reference \cite{dvir2018spectroscopy}) suppression of current through the barrier within the superconducting gap. From the supplementary note 2 of reference \cite{dvir2018spectroscopy}, one can estimate the transmission coefficient T$_{r}$ by using the WKB formula for a square barrier of width $d$ and height $U$:
\begin{equation}
T_{r}=\exp\Bigg(-\frac{2d\sqrt{2m^{*}U}}{\hbar}\Bigg),
\end{equation}
where, $ m^{*}$ is the effective mass. Taking $m^{*}$=0.54$m_{o}$\cite{cao2013two}, $U$=5.2eV and $d$ = 10nm, we obtain $T_{r}$ $\sim$ 5$\times$10$^{-46}$. These considerations unambiguously rule out any possibility of proximity effect through $\sim$ 10 nm thick hBN barrier.

We find no transport between the graphene channel and the NbSe$_2$ flake at room temperature, verifying that there is no shorting between the two.

\subsection{Mesoscopic conductance fluctuations}

In this section, we discuss the plausible origin for the observed fluctuations of $\alpha$ in Fig. 1c of the main article. More specifically, we argue that these fluctuations are mesoscopic in nature, namely universal conductance fluctuations (UCF).

UCF are a ubiquitous quantum interference phenomenon occurring in disordered mesoscopic devices at low temperatures. These reproducible fluctuations arise due to interference of different quantum mechanical trajectories.
Ideally the UCF have a magnitude of $e^{2}/h$, but in real devices the magnitude of the fluctuations depends on the competition between the coherence length and the channel length, and also diminishes with increasing temperature. In our experiment, the magnitude of the fluctuations in $R_{xx}$ is $\sim 0.1 e^{2}/h$ at 240mK, whereas there are no detectable fluctuations at 15K.

\begin{figure*}
 \includegraphics[width=\linewidth]{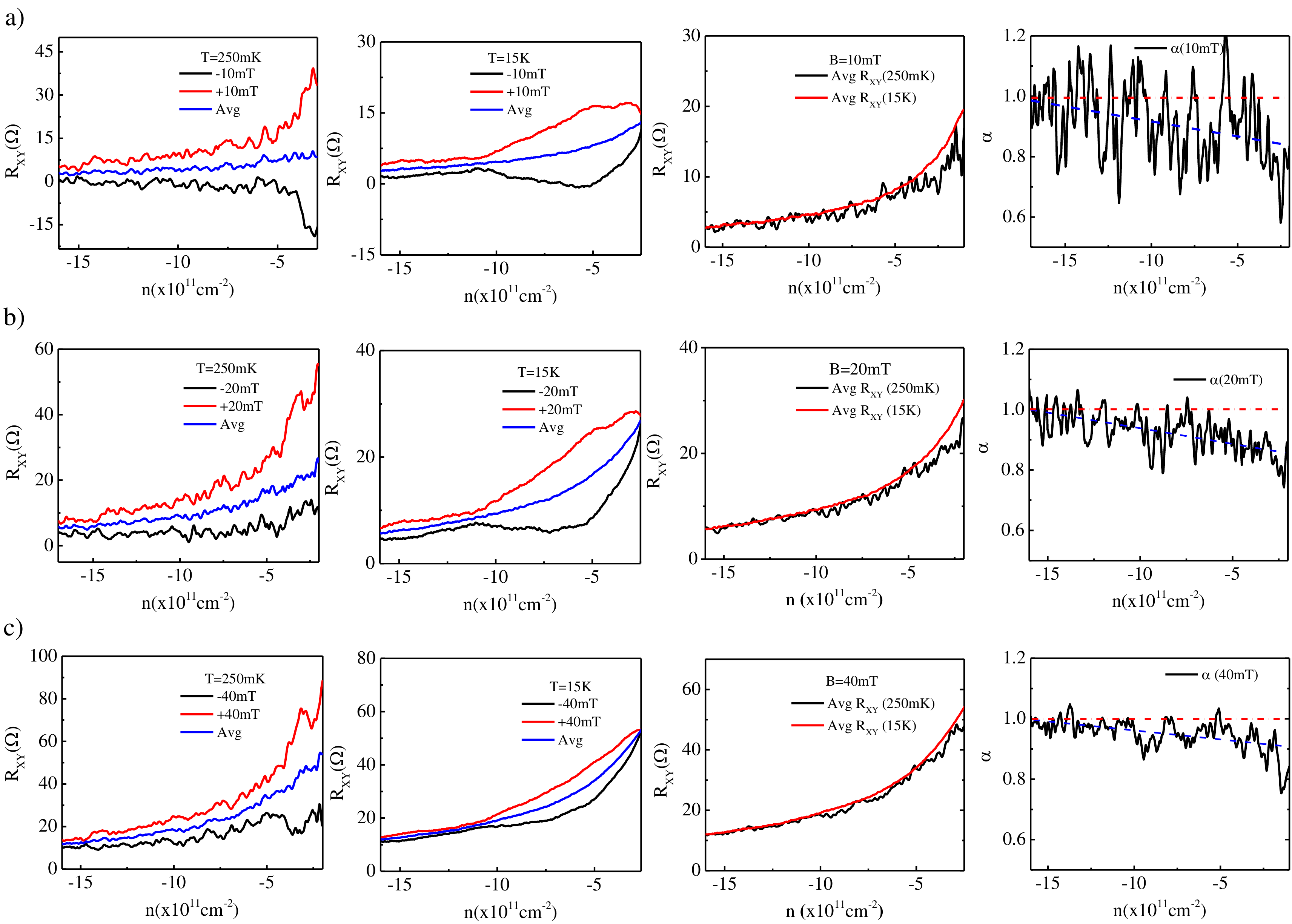}
  \caption{a) $R_{xy}$ at magnetic fields $B = +10$ mT and $B = -10$ mT, together with the average $R_{xy}$.  Two temperatures are shown: $T = 250$ mK ($T<T_{c}$) (leftmost panel) and $T = 15$ K ($T>T_{c}$) (second panel). The comparison of the average $R_{xy}$ at $B = 10$ mT for $T<T_{c}$ and $T>T_{c}$ is shown in the third panel. The parameter $\alpha = R_{xy}(T<T_c) / R_{xy}(T>T_c)$ as a function of the carrier concentration  is shown in the rightmost panel. The horizontal red dashed line corresponds to $\alpha=1$ and the blue dashed line is the linear fit used to obtain the average $\alpha$ as a function of the carrier concentration. Similarly, b) and c) show the data for $20$ mT and $40$ mT, respectively. }
  \label{expfig3}
\end{figure*}

\begin{figure*}
 \includegraphics[width=\linewidth]{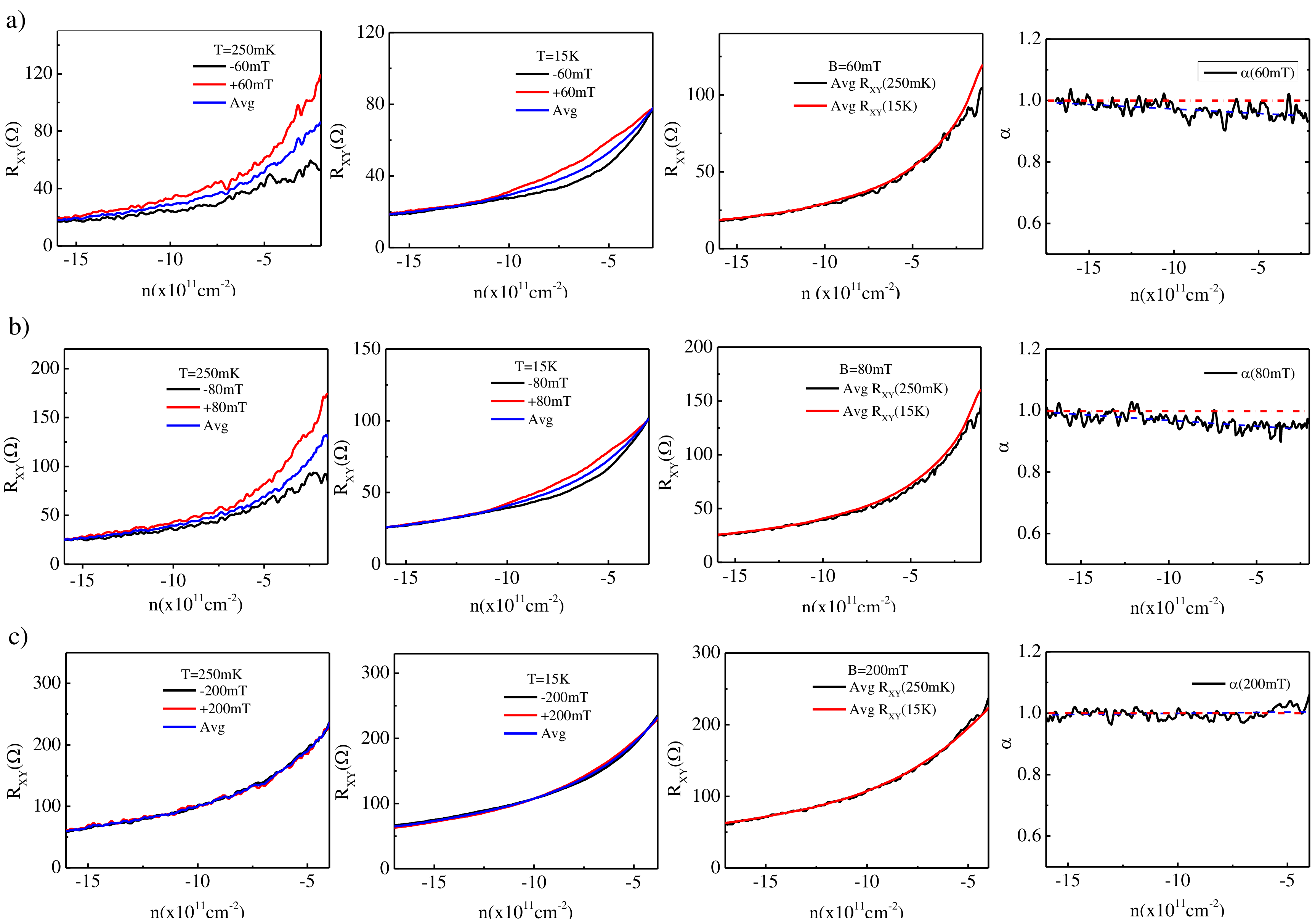}
  \caption{a) $R_{xy}$ at magnetic fields $B = +60$ mT and $B = -60$ mT, and the average value at two temperatures: $T = 250$ mK ($T<T_{c}$) (leftmost panel) and $T = 15$ K ($T>T_{c}$) (second panel). The comparison of the average $R_{xy}$ at $B = 60$ mT for $T<T_{c}$ and $T>T_{c}$  is shown in the third panel. The parameter $\alpha = R_{xy}(T<T_c) / R_{xy}(T>T_c)$ is shown as a function of the carrier concentration in the rightmost panel. The horizontal red dashed line corresponds to $\alpha=1$ and the blue dashed line is the linear fit used to obtain the average $\alpha$ as a function of the carrier concentration. Similarly, b) and c) show the data for $80$ mT and $200$ mT, respectively.  \vspace{250px}}
  \label{expfig4}
\end{figure*}

\begin{figure*}
 \includegraphics[width=1.0\linewidth]{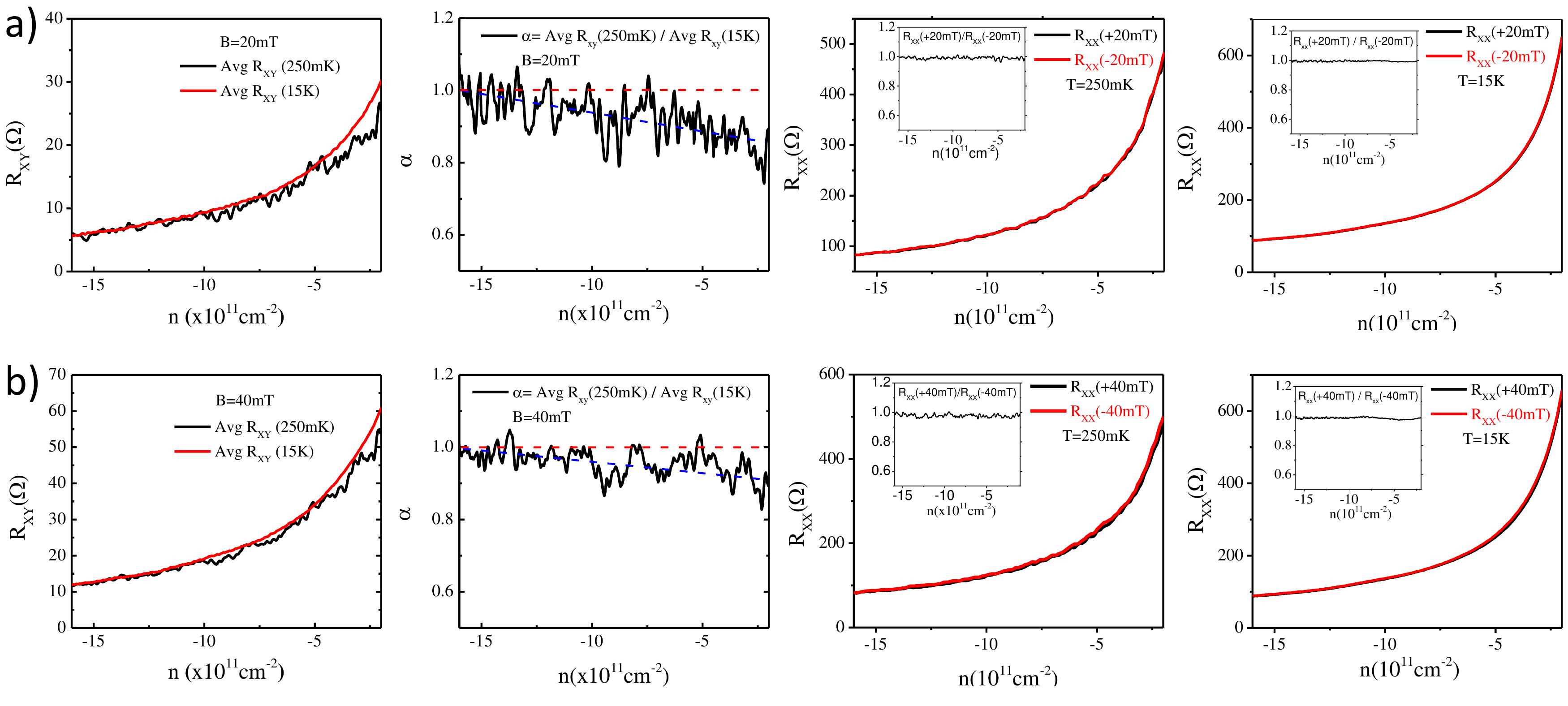}
  \caption{a) Comparison of the average $R_{xy}$ at $B = 20$ mT for $T<T_{c}$ and $T>T_{c}$ (leftmost panel). The parameter $\alpha = R_{xy}(T<T_c) / R_{xy}(T>T_c)$ as a function of the carrier concentration is shown in the second panel. The horizontal red dashed line corresponds to $\alpha=1$ and the blue dashed line is the linear fit used to obtain the average $\alpha$ as a function of the carrier concentration. $R_{xx}$ as a function of the carrier concentration for $+20mT$ and $-20mT$ at $T=250mK$ is shown in the third panel;  the inset of this panel shows the the ratio $R_{xx}(+20mT) / R_{xx}(-20mT)$ for $250mK$. The rightmost top panel shows $R_{xx}$ as a function of carrier concentration for $+20mT$ and $-20mT$ at $T=15K$; the inset shows the the ratio $R_{xx}(+20mT) / R_{xx}(-20mT)$) for $15K$. The row b) shows analogous comparison for $40mT$. }
  \label{expfig5}
\end{figure*}

We also observe the fluctuations in $R_{xy}$ (Fig.~\ref{expfig3} and Fig.~\ref{expfig4}). We attribute these to the finite $R_{xx}$ contribution. The left panels of Fig.~\ref{expfig3} show $R_{xy}(+B)$ (red), $-R_{xy}(-B)$ (black) and the average $R_{xy}=[R_{xy}(+B)-R_{xy}(-B)]/2$ (blue) as a function of density for several values of $B$. The magnitude of the UCF in the average $R_{xy}$ is much smaller than that in $R_{xy}(+B)$ or $-R_{xy}(-B)$, demonstrating that the fluctuations are reproducible to a large degree, and thus represent mesoscopic UCF rather than noise.

\pagebreak

\pagebreak